%% file: best_tech_report.tex
\documentclass[electronic]{vgtc}  

\ifpdf%
  \pdfoutput=1\relax                   %
  \usepackage{graphicx}                %
  \DeclareGraphicsExtensions{.pdf,.png,.jpg,.jpeg} %
\else%
  \usepackage{graphicx}                %
  \DeclareGraphicsExtensions{.eps}     %
\fi%

\graphicspath{{figures/}{pictures/}{images/}{./}} %

\usepackage{microtype}                 %
\PassOptionsToPackage{warn}{textcomp}  %
\usepackage{textcomp}                  %
\usepackage{mathptmx}                  %
\usepackage{times}                     %
\usepackage{cite}                      %
\usepackage{tabu}                      %
\usepackage{booktabs}                  %

\usepackage{amsmath}
\usepackage{hyperref}
\usepackage[capitalise,noabbrev]{cleveref}	%
\input{common_arXiv/cleverref-setup}

\newcommand{\mc}[1]{\mathcal{#1}} %
\newcommand{\citep}[1]{\cite{#1}} %
\newcommand{\citettwo}[2]{{#1}~\cite{#2}}
\DeclareMathAlphabet{\mathcal}{OMS}{cmsy}{m}{n}

\onlineid{1570}

\vgtccategory{Application}

\vgtcinsertpkg

\title{SPA: Verbal Interactions between Agents and Avatars in Shared Virtual Environments using Propositional Planning}

\author{Andrew Best\thanks{e-mail: best@cs.unc.edu}\\ %
        \parbox{1.4in}{\scriptsize \centering Dept. of Computer Science \\ UNC Chapel Hill}
\and Sahil Narang\thanks{e-mail: sahil@cs.unc.edu}\\ %
        \parbox{1.4in}{\scriptsize \centering Dept. of Computer Science \\ UNC Chapel Hill \\ http://gamma.cs.unc.edu/pedvr/}
\and Dinesh Manocha\thanks{e-mail: dm@cs.umd.edu}\\ %
     \parbox{1.4in}{\scriptsize \centering Dept. of Computer Science \\ University of Maryland}
}

\include{VR2020_arXiv/tex/abstract}

\CCScatlist{
    \CCScatTwelve{Computing methodologies}{Artificial intelligence}{Distributed artificial intelligence}{Multi-agent systems};
}

\begin{document}

\firstsection{Introduction}

\maketitle

\input{VR2020_arXiv/tex/teaser.tex}

\input{VR2020_arXiv/tex/introduction.tex}

\input{tech_report_vr_2020_arXiv/tex/tech_related_work}

\input{VR2020_arXiv/tex/problem_overview.tex}

\input{VR2020_arXiv/tex/planner_approach.tex}

\input{VR2020_arXiv/tex/nlp_approach.tex}

\input{VR2020_arXiv/tex/results.tex}

\input{VR2020_arXiv/tex/user_study_summary.tex}
\input{VR2020_arXiv/tex/conclusion.tex}
\input{VR2020_arXiv/tex/appendix_title.tex}
\input{VR2020_arXiv/tex/appendix.tex}

\input{VR2020_arXiv/tex/acknowledgments.tex}

\bibliographystyle{abbrv-doi}

\bibliography{common_arXiv/crowds_sahil,common_arXiv/auto_driving}
\end{document}

%% file: common_arXiv/cleverref-setup.tex
\makeatletter

\if@cref@capitalise
\crefname{channel}{Channel}{Channels}
\Crefname{channel}{Channel}{Channels}
\else
\crefname{channel}{channel}{channels}
\Crefname{channel}{Channel}{Channels}
\fi

\if@cref@capitalise
\crefname{paper}{Paper}{Papers}
\Crefname{paper}{Paper}{Papers}
\else
\crefname{paper}{paper}{papers}
\Crefname{paper}{Paper}{Papers}
\fi

\crefformat{footnote}{#2\footnotemark[#1]#3}

\makeatother

%% file: VR2020_arXiv/tex/abstract.tex
\abstract{
We present a novel approach for generating plausible verbal interactions between virtual human-like agents and user avatars in shared virtual environments.
Sense-Plan-Ask, or SPA, extends prior work in propositional planning and natural language processing to enable agents to plan with uncertain information, and leverage question and answer dialogue with other agents and avatars to obtain the needed information and complete their goals.
The agents are additionally able to respond to questions from the avatars and other agents using natural-language enabling real-time multi-agent multi-avatar communication environments.

Our algorithm can simulate tens of virtual agents at interactive rates interacting, moving, communicating, planning, and replanning.
We find that our algorithm creates a small runtime cost and enables agents to complete their goals more effectively than agents without the ability to leverage natural-language communication. 
We demonstrate quantitative results on a set of simulated benchmarks and detail the results of a preliminary user-study conducted to evaluate the plausibility of the virtual interactions generated by SPA.
Overall, we find that participants prefer SPA to prior techniques in 84\% of responses including significant benefits in terms of the plausibility of natural-language interactions and the positive impact of those interactions.
}

%% file: VR2020_arXiv/tex/teaser.tex
\begin{figure*}[h]
	\centering
	\includegraphics[width=\linewidth]{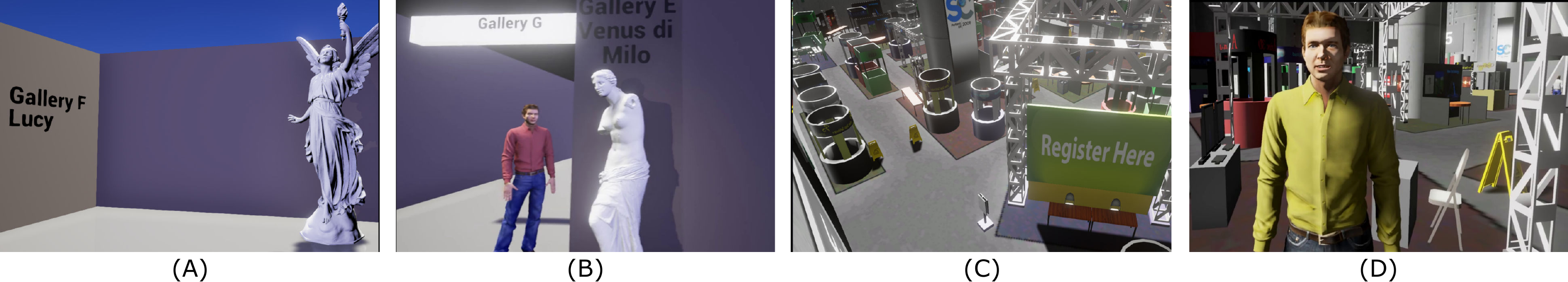}
	\caption{We performed a user-study to evaluate our novel method for generating verbal interactions between virtual agents and user-avatars. Participants in our study preferred our approach, SPA, in 84\% of responses. Participants also indicate strong preferences for our approach in terms of plausibility of interactions and how well the scenario reflected real-world scenarios. Our study consisted of two trials. \textbf{(A)}: In the museum, the avatar searches for the statue of Lucy in a gallery at the far-side of the museum.  \textbf{(B)}: While exploring, agents approach the avatar to ask the location of other statues the avatar has seen previously. The avatar can choose to provide information to the agents. \textbf{(C)}: In the tradeshow, the avatar must find the registration booth pictured. \textbf{(D)}: Using our method, the avatar can ask virtual agents for the location of the registration booth. }
	\label{nlpcrowds:fig:userstudy_benchmarks}
\end{figure*}

%% file: VR2020_arXiv/tex/introduction.tex
There is great recent interest in generating immersive social experiences.
Increasingly, games, training, and entertainment seek to provide a user with the experience of embodying a digital \textit{avatar} and sharing a virtual space with other user-controlled avatars as well as computer-controlled characters, or \textit{agents}.
Such multi-agent multi-avatar applications range from immersive games, social virtual-reality (VR) hangouts, training simulations ~\cite{rickel1999virtual,hill2003virtual}, treating social-phobias \cite{pertaub2002experiment} or visiting virtual spaces such as museums or landmarks. 

The plausibility and effectiveness of multi-avatar simulations can be improved by the presence of interactive human-like virtual agents~\cite{Garau:2005:RPV}.
However, virtual agents that do not interact in plausible ways can reduce the sense of presence in virtual environments \cite{slater:2009:experience,Bailenson:2005:IIE}.
Moreover, the context of the simulation may necessitate agents that have independent goals and are not purely focused on the co-present avatars.
The virtual world should feel like a place the avatar is visiting, as opposed to one constructed purely for the avatar.
In such cases, agents must be capable of engaging in meaningful interactions with avatars and other agents, either proactively or in response to the actions of others.
These interactions may include both verbal as well as non-verbal means of communication including  movement and navigation, gesturing, gazing etc.
Recent studies have highlighted the critical role of verbal communication and its significant impact on the perceived naturalness of user-agent interactions, and the overall effectiveness of the application~\cite{novielli2010user,nass1994computers}.

Most prior work in enabling interactions between avatars and agents is limited to embodied conversational agents (ECA), wherein an anthropomorphic virtual agent demonstrates human-like face-to-face communication~\cite{cassell2000conversation}. 
However, ECA is generally restricted to single agent-avatar pairwise interactions and is often avatar-centric. 
The agent participates in interaction with the aim of assisting the avatar in achieving a goal, or foiling the avatar, but does not plan its own intentions outside the context of the avatar-agent interaction.
There is also prior work in multi-agent navigation that has explored communication behaviors \cite{kullu2017acmics,Pelechano06}. 
However, these methods rely on message-passing or implicit communication which preclude verbal interaction with user-controlled avatars.
Overall, simulating plausible verbal interactions in shared multi-avatar multi-agent environments remains a challenge.

There are several core challenges in simulating the behaviors of virtual agents in such multi-avatar multi-agent environments. 
First, agents must be capable of independently planning egocentric behaviors in potentially uncertain conditions. 
Much like the real world, an agent may possess an imperfect understanding of the world and must be capable of proactively communicating with other entities to derive knowledge such that it can accomplish its goal.

Second, agents must be capable of communicating with avatars and other agents in unstructured conditions. 
In effect, agents must be able to interpret language, generate meaningful responses and exchange information, agnostic of whether the other entity is a user-controlled avatar or another virtual agent. 

Third, agents must be capable of generating plausible behaviors, including asking and answering questions, based on their interpretable understanding of the virtual world. 
In effect, agents should be able to absorb information through communication and behave appropriately based on the new information.

\textbf{Main Results}: In this paper, we seek to address the problem of simulating many virtual agents that can effectively plan individual actions, interact, and communicate with avatars and other agents using natural language. 
To this end, we present \emph{Sense-Plan-Ask (SPA)}, an interactive approach to enable virtual agents to accomplish their individual goals  with uncertain information in complex multi-agent multi-avatar environments (Section 3). 
The SPA approach consists of following novel contributions:
\begin{itemize}
	\item \textbf{Propositional-planning with Automatic Uncertainty Resolution}: We present a least-commitment-based planning approach~\cite{Russell:2009:AIM:1671238} to generate agent action plans with uncertain information. Agents automatically generate uncertainty resolution actions that may include navigational actions to explore the environment, or asking questions. Moreover, agents re-plan based on new information.
	\item \textbf{Multi-agent Natural language Interaction}: We present a natural language communication approach that can parse utterances received from other agents and avatars, generate natural language responses as well as construct queries and learn new information based on propositional logic. 
    \item \textbf{Proactive agents}: Our approach, SPA, allows agents not only to react to avatars and agents, but to proactively seek out interaction and engagement. The agents learn from interaction and respond accordingly, generating diverse and comprehensive simulations.
    
    \item \textbf{User Evaluation}: We present the results of a user study which demonstrates our method's advantages over prior approaches. Compared to methods which do not enable natural-language interaction between agents and avatars, participants showed significant preference for our method in terms of the plausibility of the scenarios and quality of agent-avatar interactions.
\end{itemize}

The rest of the paper is organized as follow: In \cref{nlpcrowds:sec:nlp_related}, we detail relevant related work in multi-agent systems and task planning. 
We give an overview of SPA in \cref{nlpcrowds:sec:nlp_overview}.
In \cref{nlpcrowds:sec:nlp_planner}, we detail our propositional-planning framework which enables planning under uncertainty via interaction.
\cref{nlpcrowds:sec:nlp_approach} details our natural-language processing and generation approach.
We describe our simulation benchmarks, offer performance results, and detail the results of a user evaluation of our method in \cref{nlpcrowds:sec:nlp_results}.

%% file: tech_report_vr_2020_arXiv/tex/tech_related_work.tex
\section{Related Work}
\label{nlpcrowds:sec:nlp_related}
In this section, we give an overview of relevant work in action-planning, multi-agent simulation, and verbal communication.

\subsection{Action Planning}

Action planning, sometimes referred to as classical planning, task planning, or logical planning in the literature \citep{Russell:2009:AIM:1671238, ghallab2004automated}, deals with constructing a set of actions taken by an intelligent agent to solve a problem or achieve a goal.
Classical approaches such as STRIPS \citep{fikes1971strips}, ADL, \citep{pednault1989adl} and PDDL \citep{fox2003pddl2} construct domain description formalisms which allows many distinct domains to be solvable with a common approach.
Recent work has addressed uncertainty through continual planning \citep{brenner2009continual}, or by encoding sensing actions \citep{petrick2004extending}. 
The latter approach encodes a specific ``sensing'' action for each kind of information the agent may need.
Our approach seeks to minimize explicit sensing actions, instead employing natural-language interactions and generic exploration to resolve uncertainty.
Epistemic planning approaches \citep{eger2018keeping,martens2015ceptre}, typically employed in agent-interaction scenarios, typically solve uncertainty via each agent modeling their perception of the beliefs of other agents in a turn-based interaction.
These models are inherently complimentary to our overall algorithmic approach.
However, it is unclear how an agent would determine the best question to ask given an epistemic model or how ``overhearing'' and indirect communication are modeled in these cases.

\subsection{Human and Single Agent Interactions}

Interactions between humans and automated agents can be subdivided into human-robot interactions (HRI) and human interactions with a digital or virtual agent, also called Embodied Conversational Agents (ECA). 

In the domain of HRI, robots typically plan dialog to clarify a human operator's intentions in a collaborative context~\citep{doshi2008spoken,bisk2016natural}.
The robot does not maintain independent goals or a non operator-centric representation of the domain.
The robot works as a subservient collaborator, typically limited to interaction with a single human operator.
Recent work has demonstrated a robot requesting assistance in a specific collaborative-task \citep{tellex2014asking}. 
Our framework is complimentary to the proposed language model and could be used with such a system for embodied conversation on a physical robot.
Some work has incorporated sensor uncertainty \citep{tenorth2017representations}, but modeling unknown information remains a challenge across domains.
Our approach allows agents to act on their own goals and employ natural-language interactions with multiple avatars or agents by specifying a range of domain information types without fully specifying the agent's knowledge.

Prior work in Embodied Conversation Agents (ECA) has single-agent communication with a human avatar using complex but well-structured dialogue \citep{graesser2014learning}. \citettwo{Rickel and Johnson}{rickel1999virtual} demonstrated positive effects on team training using virtual agents in a mixed agent-avatar environment. In these cases, the agent's behavior is fundamentally user-centric i.e. rather than behaving as an independent entity, the agent's plan revolves around the avatar. 
Some work has demonstrated capabilities in learning from natural language interaction \citep{Wang2016, Thomason2015a}. However, these methods limit the interactions to a single user and do not generalize to unstructured interactions. 

\subsection{Multi-agent Planning}

Action planning for interactive multi-agent systems has typically been limited to locomotion-based actions, and comprises of choosing an appropriate goal position for each agent and computing its trajectory. A large body of research exists on planning paths to an agent's goal \citep{Snook00,tollPLE} and local-avoidance behaviors \citep{ORCA,karamouzas2014universal,Schadschneider01,helbi95} for agents. Our work is complimentary to these approaches and our agents can use any of these methods for path planning.
However, many of these approaches rely on rigid finite-state machines to generate goals for the agents \citep{curtis2016menge,ulicny2002towards}. Some approaches have been proposed to incorporate contextual interactions \citep{Paris09,Shao05}, but in these cases the behaviors of the agents are pre-encoded and activated when specific conditions are met. 
Some techniques, such as \citep{yeh08}, use proxy agents to model social behaviors.
Instead, our approach allows for dynamically creating action plans for agents without a rigid or pre-encoded structure. 
Our agents generate and execute plans as needed to satisfy diverse goals.
Agents can plan simple goal positions, or complex goals, i.e. interacting with a specific agent or acquiring some item from the environment.

\subsection{Communication in Multi-agent systems}

Some prior work has introduced communication capabilities in multi-agent systems using message-passing or packet-based approaches to model interaction 
\citep{kullu2017acmics}, auditory cues \citep{huang2013spread}, social contagion \citep{chao2010simulation}, or information sharing \citep{Pelechano06}.
In each approach, agents communicate using a strict message structure, or implicitly through the sharing of data and thus preclude the use of interactive verbal communication.
\citettwo{Sun et al.}{sun2012animating} generated animated conversation for agents in a simulated crowd, and \citettwo{Brenner and Kruijff-Korbayova}{brenner2008continual} leveraged message-passing to enable multiple agents to collaborate with a simulated user (i.e. not an actively controlled avatar) in a shared-task.
In each case, the natural-language dialog was generated post-simulation as an animation feature. By contrast, our approach allows the agents to communicate with other agents and avatars using plausible verbal communication in real time.

%% file: VR2020_arXiv/tex/problem_overview.tex
\section{Background and Algorithm Overview}
\label{nlpcrowds:sec:nlp_overview}

Our approach, Sense-Plan-Ask, couples propositional planning with natural-language processing to enable many agents to interact with other agents and avatars in shared environments. This section introduces relevant terms and notation used throughout the chapter, and provides an overview  of our approach.

\begin{figure}[h]
	\centering
	\includegraphics[width=1.0\linewidth]{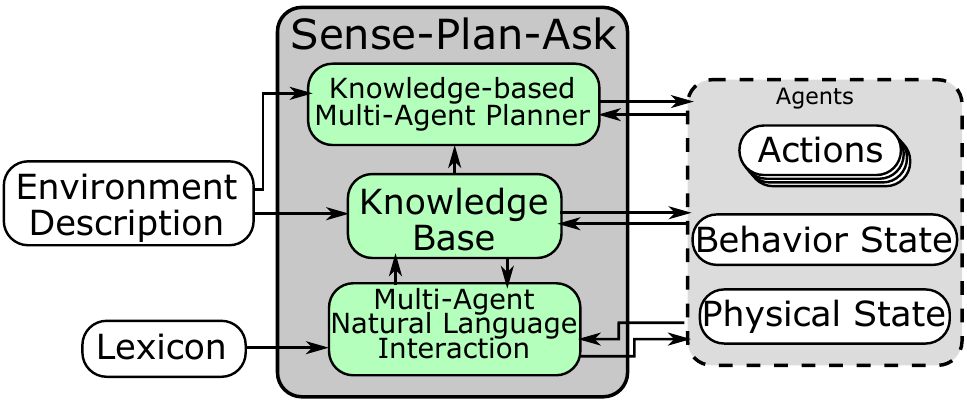} 
    
	\caption{\label{nlpcrowds:fig:pipeline} \textbf{Algorithmic Pipeline}: We present a novel interactive approach called Sense-Plan-Ask (SPA) which enables agents to interact with other agents and user-controlled avatars using natural language communication. Each agent accomplishes individual goals using SPA to compute explainable action plans based on uncertain information. The agent engages in natural language conversations with nearby agents and avatars to resolve uncertainty in its plan. The communication model increases the plausibility of the simulation and improves the user's experience in immersive virtual environments.
	}
\end{figure}

\subsection{Agent Model}

Each agent in the virtual environment can be defined by its \emph{physical state} and its \emph{behavioral state}. The physical state comprises of physical properties such as position, velocity and its current action. The \emph{behavior state} regulates its decision making, including its current knowledge and the set of available actions.

\subsubsection{Physical State}

For the purpose of behavior and movement planning, we treat each agent $i$ as a bounded disc in a 2D plane with scalar radius $r_i$. 
We denote the agent's position and velocity at time $t$ by $\vec{p}_i^t$, and $\vec{v}_i^t$ respectively. 
Furthermore, we denote the current action being executed by the agent as $a_{i}^t \vert a_{i}^t \in \mathcal{A}_i$.
$\mathcal{A}_i$ is the set of all actions available to the agent and may include movement or navigation as well as non-movement actions, such as speaking, or interacting with the environment, i.e. affecting change to the simulation domain.
We assume a path planner is available for locomotion and is used to compute the path and avoid collisions with other agents, avatars, and obstacles in the environment.
Collectively, the position, velocity, and current action describe the agent's time-varying physical state $\{ \vec{p}_i^t, \vec{v}_i^t, a_i^t\}$.

\subsubsection{Behavior State}

We conceptualize our agents as Belief Desire Intention (BDI) agents \cite{Rao1995}. Each agent maintains an independent understanding of the world represented by a set $\mathcal{B}$ of facts called \emph{beliefs}.
We use the term belief to reflect the fact that the agent's knowledge may be wrong.
Moreover, each agent is given a set $\mathcal{D}$ of high-level \emph{desire}s, or goals, to achieve in the simulation.
Each agent plans a set of intermediate actions called \emph{intentions} to accomplish these desires.

We encode the agent's beliefs, desires, and intentions in a first-order propositional planning language and store the agent's knowledge in relational database. %
Our formalism is a subset of ADL \cite{pednault1989adl} with the exception that we do not allow disjunction in desires, and limit desire specification to grounded literals. 
First-order propositional planning provides sufficient representational power for our interactive benchmarks (see \cref{nlpcrowds:sec:nlp_results}) and fits well with the types of questions we support in our interaction.
We define a set $\Sigma$ comprising of \emph{entities} that are symbolic constants through the simulation. These may include agents, items, physical locations, etc. We apply a \emph{type} to each entity and a set of attributes associated with the type. 

We describe predicates as first-order formulae over $\Sigma$. We allow constraints on the type of arguments in the predicate schema to reduce the search space, and we explicitly allow negated predicates in state specification. Furthermore, we categorize predicates in two categories. \emph{Knowledge predicates} describe relationships or facts the agent might know. \emph{Fluent predicates} describe transitive properties the agent may hold, i.e. being at a specific location \cite{Russell:2009:AIM:1671238}. A belief-state for agent $i$ at time $t$, denoted $\mc{B}_i^t$, consists of the set of all beliefs known to be true or false to the agent. For compactness, we assume $\Sigma$ is known to all agents and is implicitly included in $\mc{B}$. Consider a problem-solving domain consisting of a series of keys and locks, the following would be a valid state specification:

\begin{align*}
    \mc{B}_i^t = \{&Have(Key_1), \neg Have(Key_2), Opens(Key_1, Safe_1),\\ 
              &\neg Locked(Safe_2)\}
\end{align*}

We do not assume a complete state specification. That is, $B_i^t$ contains all \emph{known} information. Missing predicates are considered unknown as opposed to false. In the example above, agent $i$ knows that $Key_1$ opens $Safe_1$, but does not know of anything which opens $Safe_2$. In \cref{nlpcrowds:sec:nlp_planner}, we detail how our planning approach allows agents to plan uncertainty resolution and in \cref{nlpcrowds:sec:nlp_approach} we describe how they use verbal communication to resolve uncertainty.

We define a set of operators, or \emph{actions} $\mc{A}$, over beliefs of the form $O<parameters, conditions, effects>$. Parameters describe elements in $\Sigma$ passed to the action, subject to type constraints on the element. Conditions are predicates which must hold in $\mc{B}_i^t$ for the action to be applicable. Effects are predicates added to $\mc{B}_i^t$ upon application of the action. Should a predicate in $\mc{B}_i^t$ be contradicted by new information, it is removed 
corresponding to the agent updating its beliefs. 
 Continuing from the earlier example, the following would be a valid action schema corresponding to opening the safe: 

\begin{align*}
    Open<&\{key: X, safe: Y\}, \\
         &\{Locked(Y), Opens(X,Y), Have(X)\},\\
         &\{\neg Locked(Y)\}>
\end{align*}

The desires of the agent are a time-varying set $\mc{D}_i^t$ of beliefs which must be achieved by actions over the agent's initial state, $\mc{B}_i^0$. The agent's behavior state for time $t$ is compactly described as $\{\mc{B}_i^t, \mc{D}_i^t, \mc{A}_i^t\}$.

\subsection{Sense-Plan-Ask (SPA) Algorithm}

Our proposed approach, Sense-Plan-Ask, generates plausible interactions between virtual agents by coupling a novel propositional 
planner with a natural-language processing framework. SPA is an agent-based simulation algorithm 
which enables the agents to communicate, plan, and interact with other agents and avatars. \Cref{nlpcrowds:fig:pipeline} details our algorithmic pipeline.

\textbf{Sense}: We conceptualize the simulator as a discrete in time and continuous in space. The simulation updates at a fixed rate, $\Delta t$. Each update, the agents in the simulator ``sense'' their surroundings. They observe relevant predicates associated with nearby entities within a range, and ``hear'' utterances produced by agents and avatars in their vicinity. Based on the observations, the agents update their internal knowledge representation and react accordingly.

\textbf{Plan}: Each agent plans a series of actions to accomplish its goals. \Cref{nlpcrowds:sec:nlp_planner} describes how these plans are constructed based on uncertain information, and how the planner generates disambiguation actions such as asking questions of nearby agents or avatars to resolve the uncertainty. The planner allows the agents to process new information and re-plan rapidly, as is detailed in \cref{nlpcrowds:sec:nlp_results}.

\textbf{Ask}: The natural-language approach combines shallow semantic parsing with template-based generation to produce intelligible verbal questions and answers between agents and avatars. \cref{nlpcrowds:sec:nlp_approach} details how agents learn new information from their interactions and use that information to resolve uncertainties in their plans. In addition, agents can interact with avatars and other agents verbally by engaging in question and answer-based dialogue. 

Each simulation environment is configured from a domain specification file, which contains the set of entities, predicate schema, action schema, and initial knowledge for each agent in the environment. This specification also gives initial conditions of agents and other objects in the environment. This domain description is paired with an English lexicon to automatically generate training data for a shallow-semantic parser as described in \cref{nlpcrowds:sec:nlp_approach}. 
This domain specification is also used to construct a query-able knowledge-base, encoded in a SQL database, for each agent which allows the agents to recall and respond to changes in the environment.

In the following sections, we describe our algorithms for linking propositional planning and natural-language parsing and generation. The SPA approach is general, however, and could be extended to other planning approaches or natural-language generation methods with some necessary adaptation of the translation mechanism from uncertainty to natural-language utterance.

%% file: VR2020_arXiv/tex/planner_approach.tex
\section{Knowledge-based Multi-Agent Planning with Incomplete Information}
\label{nlpcrowds:sec:nlp_planner}

Our proposed planning algorithm enables each agent to plan actions to achieve its desires based on its current, potentially incomplete set of beliefs. Our algorithm tracks the uncertainty in the agent's plan and determines new actions such as verbal interaction or exploration of the environment to resolve the uncertainty. Moreover, it enables agents to update their beliefs in real-time and re-plan based on their new set of beliefs. 

\subsection{Two-stage Action Planning with Incomplete Information}

In many propositional planning languages, predicates absent in the state description are considered false, and uncertainty is prohibited \cite{ghallab2004automated}. Typical first-order propositional planning approaches may allow incomplete specifications, but require an explicit action for determining the truth state of each individual pre-condition of an action.
This can lead to a combinatorial explosion in the number of actions and subsequently the planning time of the approach. \cite{Russell:2009:AIM:1671238}.  Our approach addresses these limitations and allows for efficient planning under uncertainty by leveraging a two-stage planning algorithm.

Given agent $i$'s desire set, $\mc{D}_i$, a set of actions must be constructed to satisfy all of the desires. We apply a least-commitment, backward state-space search approach \cite{Russell:2009:AIM:1671238}. Each planning step may comprise of multiple iterations. In each iteration $j$, the agent chooses an action which satisfies the first desire $d_0 \in \mathcal{D}_i^j$. A new desire set $\mathcal{D}_i^{j+1}$ is created consisting of the remaining unsatisfied desires and any unsatisfied pre-conditions of the chosen action. 

As described in \cref{nlpcrowds:sec:nlp_overview}, we differentiate between knowledge and fluent predicates when computing $\mc{D}_i^{j+1}$. Knowledge predicates present in $\mc{B}_i$ are considered satisfied, and any absent from $\mc{B}_i$ are added to an uncertainty set, $U_i$, as opposed to the new desire set, $\mc{D}_i^{j+1}$. Any other arguments of the action aside from those needed to satisfy the desire are left unbound, i.e. they are not assigned any entity. For each unbound argument $k$, a candidate set of all entities which may satisfy $k$ is constructed, termed $C_k$. Candidates are chosen based on two criteria: they must satisfy any type and property constraints of the predicate, and, given the candidate binding, all pre-conditions of the action must be either hold in $B_i$ (true) or be absent from $B_i$ (unknown).

Our planning algorithm continues in this fashion, achieving desires until either the desire set is empty, $D_i^j = {\emptyset}$, or the desire set represents a subset of the agent's current belief state, $D_i^j \subset B_i$.
The result is a plan-template, $\mc{P} = \{A_i^{t_0}, ..., A_i^{t_n}\}$, i.e. a sequence of actions which satisfy the agent's desires, a set of candidates for all unbound arguments $\mc{C} = \{C_k | k \in K\}$, and a set of unknown predicates,$U_i$,  associated with the plan. The action order is inverted for execution, consistent with the backward state-space search.

The second stage of our planning algorithm finds appropriate bindings for the candidates in $\mc{C}$.
We first construct a set of grounded plans such that all candidates are given a specific binding. 
These plans are sorted according to the number of predicates from $U_i$ found in $B_i$ given the candidate bindings. 
In effect, the agent prefers plans with the least uncertainty. 
For each predicate in $U_i$ not found in $B_i$, an uncertainty resolution action is inserted into the plan prior to the first occurrence of the predicate.
Uncertainty resolution actions include exploring the environment or asking questions.
The final plan now consists of the original actions in $\mc{P}$ and an uncertainty resolution for each unknown predicate.
We refer the reader to the supplemental material, available at our URL, for a diagram of our planning approach.

\subsection{Plan Execution and Re-planning}

Each action described in the problem domain is mapped to a simulation controller in the agent. The controller is responsible for executing the action in the simulation environment. These controllers include uttering, moving to a location, interacting with the environment, waiting a specific amount of time, etc.

If a controller fails to accomplish the specified action for the agent, or the agent is unable to acquire the the information needed for an uncertain belief, the plan binding fails. If other bindings are available for all entities in $\mc{C}$, the next binding in order of uncertainty is chosen. Each time an agent acquires new information, the uncertainty of remaining bindings is updated and the set of candidates adjusted to prevent repetitive questions. 

If no suitable bindings are available, the planner discards the plan template and back-tracks to the prior branch in the backward state-space search. If no additional branches can be chosen, the plan is discarded and the planner fails. The agent waits a pre-defined amount of time before restarting the planning procedure.
These failures are often caused by a failure to acquire information, and are likely to succeed on subsequent planning attempts as other agents and avatars move near the agent.

%% file: VR2020_arXiv/tex/nlp_approach.tex
\begin{table*}[]
	\centering
	\caption{Sample mapping from the shallow parser to knowledge queries in the museum benchmark (\cref{nlpcrowds:sec:nlp_results}). Utterances are parsed into NL-Is (natural-language intentions) and NLEs (natural language entities). Entities are matched to relationships and a knowledge query is constructed.}
	\label{nlpcrowds:table:utterence_responses}
	\begin{tabular}{|p{3cm}|p{3cm}|p{5cm}|p{4cm}|}
		\hline
		Utterance & NL-I & NLEs &  Mapped Belief \\ \hline
		where is the & predicate question & knowledge entity: venus de milo & InSpace(?,Venus) \\
		venus de milo & & predicate: where &  \\ \hline
		
		What material is & attribute question & knowledge attribute: material & statue(Venus).material=? \\
		the venus de milo &  & knowledge entity: venus de milo & \\ \hline
		
		venus de milo is & predicate answer & knowledge entity: venus de milo & InSpace(Venus, GalleryA) \\
		located in gallery a & &predicate: located & \\
		& & knowledge entity: gallery a & \\ \hline
	\end{tabular}
\end{table*}

\begin{figure*}[]
	\centering
	\includegraphics[width=\textwidth]{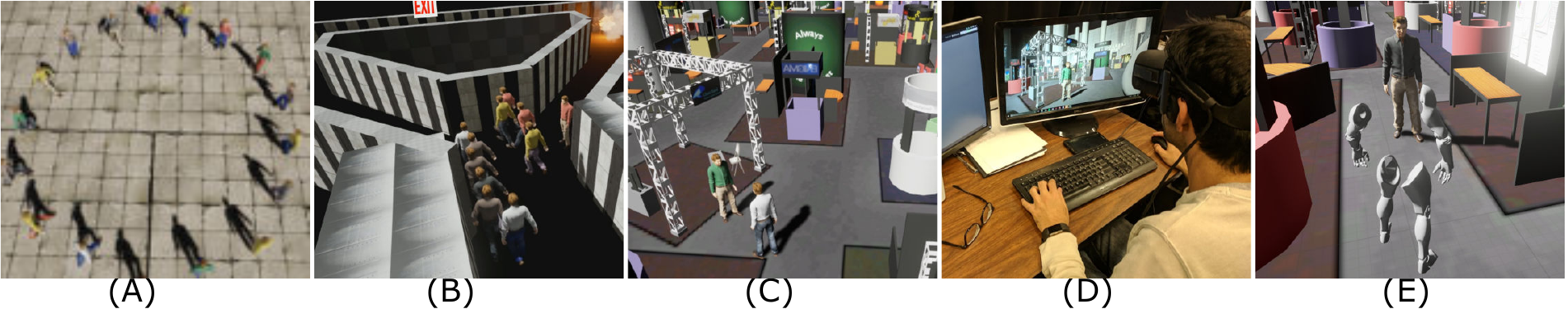}
	\caption{\textbf{Simulation Benchmarks}: We demonstrate our algorithm's performance in a set of simulated environments. \textbf{ (A)} Agents in the anti-podal circle scenario search for items while requesting information from one another. \textbf{ (B)} Agents evacuate a building during an emergency. One agent acts as a first-responder, warning the other agents to avoid the blocked hallway. \textbf{ (C)} Agents explore a densely-packed tradeshow scene, visiting booths. Agents request location and booth information from one-another to resolve uncertainty and reach their goals more effectively. \textbf{ (D)}: A user explores the tradeshow with a first-person view in immersive settings. Our approach allows virtual agents to interact with both agents and avatars simultaneously. \textbf{ (E)}: The user's avatar (shown from a third-person view) interacts with a virtual agent in the tradeshow.}
	
	\label{nlpcrowds:fig:benchmarks}
	\vspace*{-0.1in}
\end{figure*}

\section{Natural Language Interaction between Virtual Agents and Avatars}
\label{nlpcrowds:sec:nlp_approach}

This section details our natural-language processing and generation approach that allows virtual agents to engage in natural-language interactions with other agents and user-controlled avatars. 
This includes responding to agent or avatar questions and statements, as well as posing questions which resolve uncertainty in the agent's plans and facilitate achievement of the agent's goals.

\subsection{Parsing Natural Language Utterances}

To understand incoming utterances, our agents leverage state of the art shallow semantic parsing \cite{rasanlu}. 
Shallow semantic parsers are trained on a corpus of example sentences to label the sentence with an ``intention'' and extract from it natural-language \emph{named-entities}~\cite{manning2014stanford}. 
To avoid confusion with knowledge entities, we refer to named-entities as NLE (natural-language entity) and intentions as NL-I (natural-language intention).

The NL-I of an utterance is a label which is used to categorize the utterance and determine an appropriate response. 
We provide training examples for specific questions and answers our agents should be capable of responding to. 
We refer to these as in-domain NL-I. 
They are: predicate question, predicate answer, attribute question, and attribute answer.
Aside from domain specific NL-I, we provide example data for five generic NL-I, which we refer to as out-of-domain NL-I. 
These are: greeting, thanks, farewell, affirmation, and fallback (i.e. random dialogue). 
 \Cref{nlpcrowds:sec:training_parser} details how we generate training sentences automatically for our in-domain NL-I.

NLEs are specific words in an utterance which are considered important for understanding its meaning.
Typically, example sentences for each NL-I are also annotated with the relevant NLEs. 
We specifically provide training data for five NLEs: attribute instances, attribute types, predicate types, knowledge entities, and addressees. 
\Cref{nlpcrowds:sec:training_parser} details how we generate training sentences automatically for our target NLEs.

As an example, the utterance ``Where is object A?'' would receive the NL-I \emph{predicate question} as it relates to the location of the object. 
The parser would also recognize two NLEs, ``object A'' of type \emph{knowledge entity} and ``where'' of type \emph{predicate instance}.

\subsubsection{Training the parser}
\label{nlpcrowds:sec:training_parser}

We generate training data by coupling our domain descriptions with an English lexicon.  
The lexicon provides part of speech information for the set of words in our problem-domain as well as subjective and objective verb tense information. 
The lexicon also provides a set of sample usage sentences which we annotate with NL-Is and NLEs.
To train the parser, sample sentences are drawn from the lexicon and the template parameters are bound to corresponding entries from the knowledge base. 
We also provide a set of basic responses for the set of out-of-domain NL-Is.
Limiting the shallow parser to few NL-Is and NLEs allows us to train the parser using tens of examples rather than hundreds or thousands used for modern voice assistants.

For the results demonstrated in \cref{nlpcrowds:subsec:benchmarks}, we created a custom, limited lexicon. 
However, our method would generalize to a common lexicon provided that the sentence templates could be extracted. Recent work~\cite{lindsay2017framer} has demonstrated the ability to extract planning domains automatically from text and may provide a potential avenue for automatic tagging of lexical entries.

\subsection{Understanding Utterances from Avatars and Other Agents}

Each agent ``hears'' utterances issued by other avatars and agents that are visible with respect to obstacles and are within a tunable hearing range. 
Each utterance is parsed and the NL-I and NLEs are returned.
For out-of-domain NL-I, the agent responds with one of the example responses provided in the lexicon.

To map an utterance to the planning framework, the recognized NL-I must be an in-domain NL-I and the utterance must contain at-least one NLE of type \emph{predicate instance}, \emph{predicate type}, or \emph{knowledge attribute}.
Each entity in $\Sigma$ is required to have a matching NLE in the lexicon.
The agent maps the recognized NLEs to their knowledge-base equivalents.
The agent constructs a belief from the entities according to whether a predicate or attribute was detected.
For predicates, entities are matched to the slots of the predicate.
For attributes, entities are mapped to the attribute relationship.

Consider this statement generated from the example above, ``Key one opens safe one''. The NLE ``opens'' would map to knowledge \emph{predicate type}.
``Key one'' and ``Safe one'' would map to \emph{knowledge entities}.
The planner would construct the complete predicate instance $Opens(Key_1, Safe_1)$.

In the case of questions, if a slot is missing from the constructed predicate or attribute belief, the missing information is used as the subject of the question. 
If no information is missing and a complete belief can be constructed, the question is assumed to be a confirmation question. 
The question ``Does key one open safe one?'' would receive the NL-I predicate\_question and same NLEs as the statement form.
The NLE ``opens'' would map to knowledge \emph{predicate type}.
``Key one'' and ``Safe one'' would map to knowledge \emph{knowledge entities}. 
In this case, the complete predicate $Opens(Key_1, Safe_1)$ would be interpreted as a confirmation question. 
Similarly, the question ``Which key opens safe one'' would map to the predicate $Opens(?, Safe_1)$ and be interpreted as a question.

In some cases, such as the question "Is object A in location B?", no specific predicate information is given. 
However, if the agent can find a predicate which accepts all the detected entities, it can be inferred from the utterance.
\Cref{nlpcrowds:table:utterence_responses} provides several example mappings for NL-Is, NLEs, and constructed beliefs from the museum benchmark (see \cref{nlpcrowds:sec:nlp_results}). 

Once a belief is constructed, the agent queries its knowledge base to determine an appropriate response to the question. 
For confirmation questions, if the agent finds a belief matching the query, the agent responds in affirmation or negation, i.e. ``Yes, Key one opens safe one.''
For information questions, the agent will issue a response for each candidate found which satisfies the query belief, i.e. ``Key one opens safe one. The master key opens safe one.''
For utterances labeled as answers, if a complete belief is constructed, it is added to the agent's knowledge base.

\subsection{Generating Questions and Statements}

\textbf{Questions:} Each uncertain predicate in the agent's plan must be resolved in order to satisfy the agent's desires.
To generate a question for an uncertain item, the agent first maps the attribute or predicate in question to its matching entry in the lexicon to discover potential template questions for the given item. 
The production templates are augmented with appropriate slots for entities, predicates, etc. 
The agent binds the entities from its uncertain predicate to the NLE slots in the production template.
If all entities in the question are bound and all NLE slots in the production are complete, the production sentence can be uttered. 
We maintain several production templates for each predicate and attribute to generate variation in the agents' utterances. 
The agent may optionally determine to whom the question is addressed.
If there is one nearby agent, the agent can specifically address the other agent using an arbitrarily assigned, unique phonetic name. 

\textbf{Statements:} Similar to questions, statements are bound by matching the knowledge predicate to a sample production template. 
Once a question is received, the agent performs a query into the knowledge-base. 
If an answer is found, i.e. a predicate or attribute belief which satisfies the query, the agent matches the belief into production templates for the attribute or predicate.
If no answer is found, the agents are given a set of generic responses representing a lack of knowledge, e.g. ``I'm sorry. I don't know.''

The agents process and produce natural language utterances as needed to respond to questions or pro-actively seek information. No implicit information is transmitted between agents, which allows the agents to communicate with agents or avatars without distinction between the methods of interaction.
The supplemental material provides an example exchange between two agents.

%% file: VR2020_arXiv/tex/results.tex
\section{Results}
\label{nlpcrowds:sec:nlp_results}

In this section, we highlight the effectiveness of our novel approach in generating interactions among virtual agents and between virtual agents and avatars given uncertain information, as well as the role of verbal communication in resolving uncertainty. For implementation and platform details, we refer the reader to the supplimental material.

\begin{table*}[th]
	\centering
	\caption{{\bf Benchmark comparisons with and without NL-I}. This table compares performance of our algorithm with  and without NL-I. We find that agents who are able to communicate accomplish their goals more quickly and benefit from nearby communication of other agents.}
	\label{nlpcrowds:table:benchmark_results}
	\begin{tabular}{llrrrrr}
		\toprule
		Scene &  Agents &  Planning Time &  Replans &  Replan Time (S) & Solution Time(S)\\
		\midrule
		Anti-podal Circle & 10 & 76.390 & 4 &  0.007006 &  67.85 \\ 
		\ \ Without NL-I&  10 &  75.917 & 4 &  0.005508 &  73.95 \\ \hline
		\hline
		Evacuation & 11 &   1.009 &  10 &   0.000278 &    36.60 \\
		\ \ Without NL-I&  11 &  1.771 &   10 &   0.000263 &  53.70 \\ \hline
		Museum  &   5 &  6.811 &  9 &   0.013800 &   159.35 \\
		\ \ Without NL-I &   5 &   12.361 &  2 &   0.006314 &   209.40 \\ \hline
		Trade Show  &   4 &  0.123 &  1 &   0.001375 &  52.65 \\
		\ \ Without NL-I &   4 &  0.128 &  1 &   0.001116 &  98.95 \\
		\bottomrule
	\end{tabular}
\end{table*}

\begin{figure*}[t]
	\centering
	\begin{tabular}{cc}
		\includegraphics[width=0.42\textwidth]{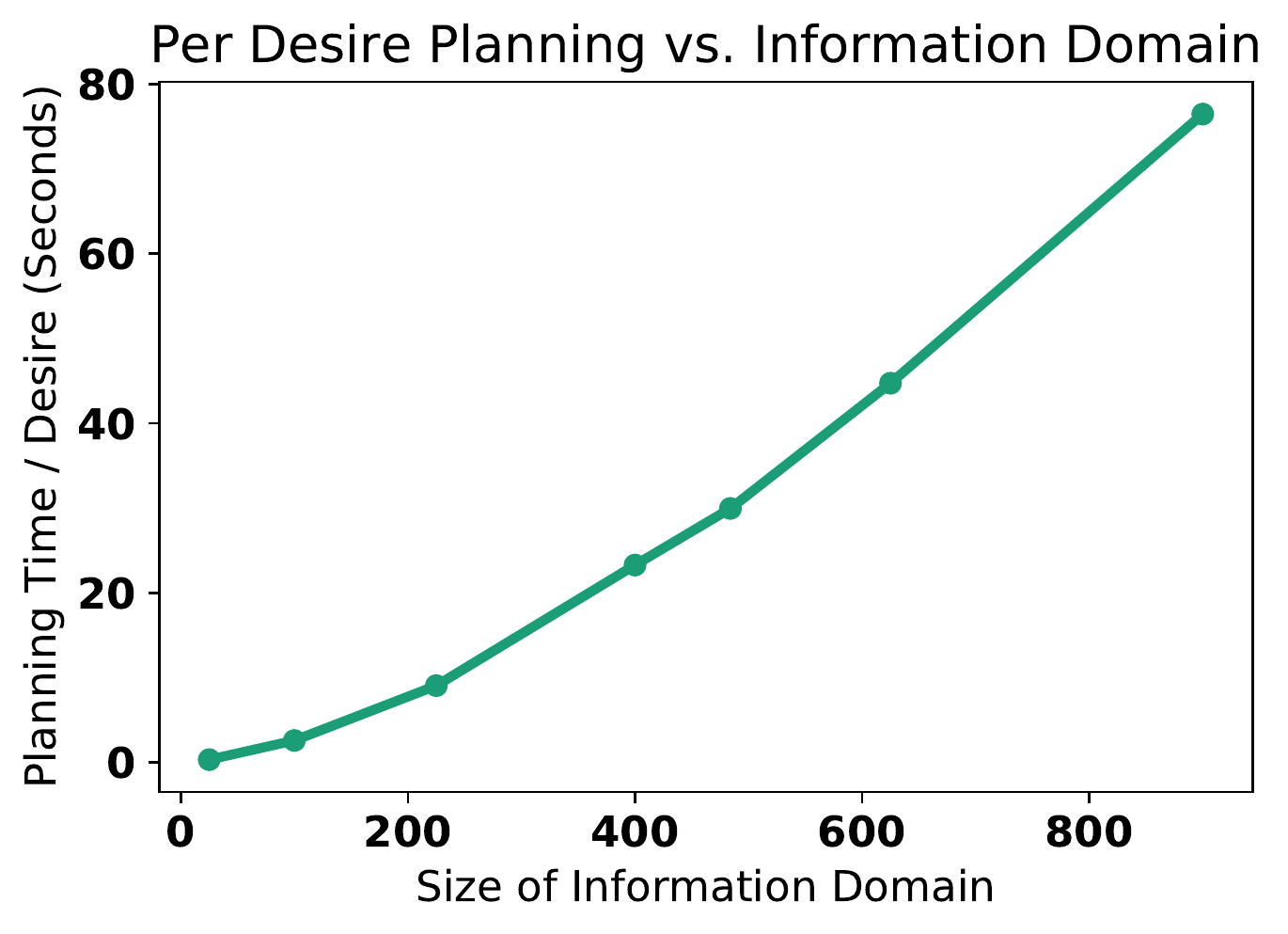} &
		\includegraphics[width=0.48\textwidth]{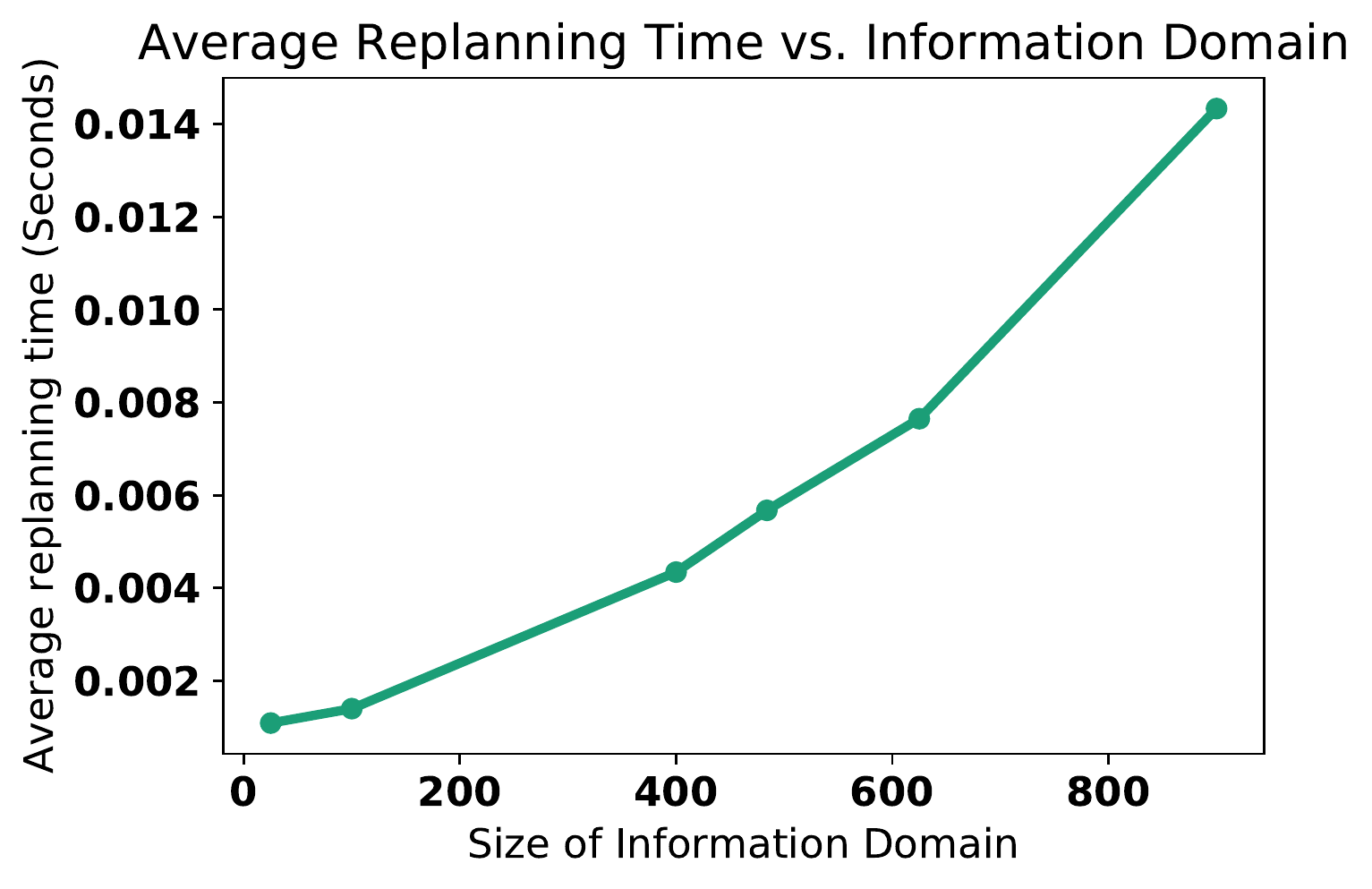} \\
		(A) & (B) 
	\end{tabular}
	\caption{\textbf{Performance Results}: \textbf{(A) Per desire planning time varying domain size:} The per-desire planning time scales exponentially as a function of domain size. This is consistent with proposition approaches. \textbf{(B) Average time per replanning:} Using our two-stage algorithm, replanning can be performed on the order of milliseconds even in complex information domains, substantially reducing planning time.}
	\label{nlpcrowds:fig:performance_micro_graphs}
	\vspace*{-0.15in}
\end{figure*}

\subsection{Benchmarks}
\label{nlpcrowds:subsec:benchmarks}
We demonstrate the results of our approach on several challenging multi-agent scenarios (\cref{nlpcrowds:fig:benchmarks}). 
\Cref{nlpcrowds:table:benchmark_results} details a performance comparison between our method and a prior crowd simulation approach without natural language communication. Overall we find that our approach decreases the overall solution time with a small increase in replanning times.
Specific details on the number of agents and desires in each benchmark can be found in the supplemental material.
The simulation benchmarks tested include:

\textbf{Anti-podal Circle:} In this scenario, 10 agents and an equal number of `goal-objects' are distributed on the circumference of a circle (\cref{nlpcrowds:fig:benchmarks}(a)). Each agent is given a desire to retrieve a set of randomly assigned goal-objects. However, the agent's initial belief set may not include information about the desired goal-objects. The agent engages in verbal communication with other agents in order to resolve uncertainties regarding the location of the goal-object. This benchmark illustrates the ability of our algorithm to plan with incomplete information, automatically generate uncertainty resolution actions, and facilitate verbal agent-agent communication. 
We find that asking questions reduced overall simulation time compared to our method without the ability to resolve uncertainty via interaction.

\textbf{Evacuation:} We simulate a fire evacuation scenario in a Y-shaped corridor (\cref{nlpcrowds:fig:benchmarks} (b)). A group of 10 agents approach a junction and must choose one of two passageways to reach their goals. Each agent randomly selects its desired passageway. 
Unknown to the agents, the right passageway is obstructed due to a fire breakout.
As the agents approach the junction, an agent acting as a first-responder redirects them to the leftmost passageway using verbal communication. 
Without our algorithm, agents are unable to communicate and some take the rightmost passageaway forcing them to retreat.
This leads to a 47\% increase in evacuation time as is shown in \cref{nlpcrowds:table:benchmark_results}.

\textbf{Avatar \& Multi-agent Tradeshow:} In this scenario, agents explore a complex tradeshow scenario with a large number of booths (\cref{nlpcrowds:fig:benchmarks} (c)). One of the agents is given the desire to go to the 'registration desk' but does not know its location. Using our two-stage planning approach, the agent generates a plan template and a set of candidates for the location of the registration desk. It then verbally interacts with nearby agents to resolve the uncertainty and find the location of the registration desk. We also simulate this scenario with a user in immersive settings (\cref{nlpcrowds:fig:benchmarks} (d)). The user controls a virtual avatar and is given the task of finding the registration desk. The user asks questions, and receives meaningful responses from the agents (\cref{nlpcrowds:fig:benchmarks} (e)).

\textbf{Avatar and Multi-agent Museum:} This scenario demonstrates an art museum. In the multi-agent case, each agent is given a set of statues to visit. However, knowledge of the locations of the statues is randomly assigned amongst the agents. Each agent must seek out other agents with whom they can interact to acquire the location of the statuary they are seeking. In the avatar case,  the user is given a specific statue to find in the museum. During the user's exploration, virtual agents approach the user and request the locations of statues the user has previously visited. If the user responds with the appropriate information, the agents are able to complete their plans.

\textbf{Multi-Avatar Hide-and-Seek:} This scenario depicts multiple user avatars engaged in a hide-and-seek game in a virtual environment populated by many virtual agents. The hiding avatar chooses a room in which to hide, and the seeking avatar must interact with the virtual agents to obtain the location of the the hider. The necessary information is disbursed amongst several agents, requiring the seeker to interact with multiple virtual agents to find the hider.

\subsection{Performance Analysis}

We conducted a series of repeated trials on the Anti-podal circle benchmark described in \cref{nlpcrowds:subsec:benchmarks}. In each trial, we increased the number of virtual agents or the number of potential goal-objects for the agents.
Overall, we find that our planning algorithm scales linearly in the number of potential targets for a single agent and linearly in the number of agents. 
Consistent with other propositional planning approaches, we find our algorithm scales exponentially in the size of the information domain. 
We further evaluated the replanning time for each agent as the agents resolve uncertainty in their plans.
We find that our two-stage approach yields negligible replanning times even though agents must resolve new information and choose new candidates. 
\Cref{nlpcrowds:fig:performance_micro_graphs} details how our two-stage approach reduces overall computation time be enabling rapid replanning.
Additional details can be found in the supplemental material.

In a typical scenario, the generation of candidates is only performed once, and can done as a pre-processing step for the scenario, leading to interactive agents capable of replanning as a response to verbal communication with extremely small overhead.

%% file: VR2020_arXiv/tex/user_study_summary.tex
\subsection{User Evaluation}
\begin{figure*}[ht]
	\centering
	\includegraphics[height=1.8in]{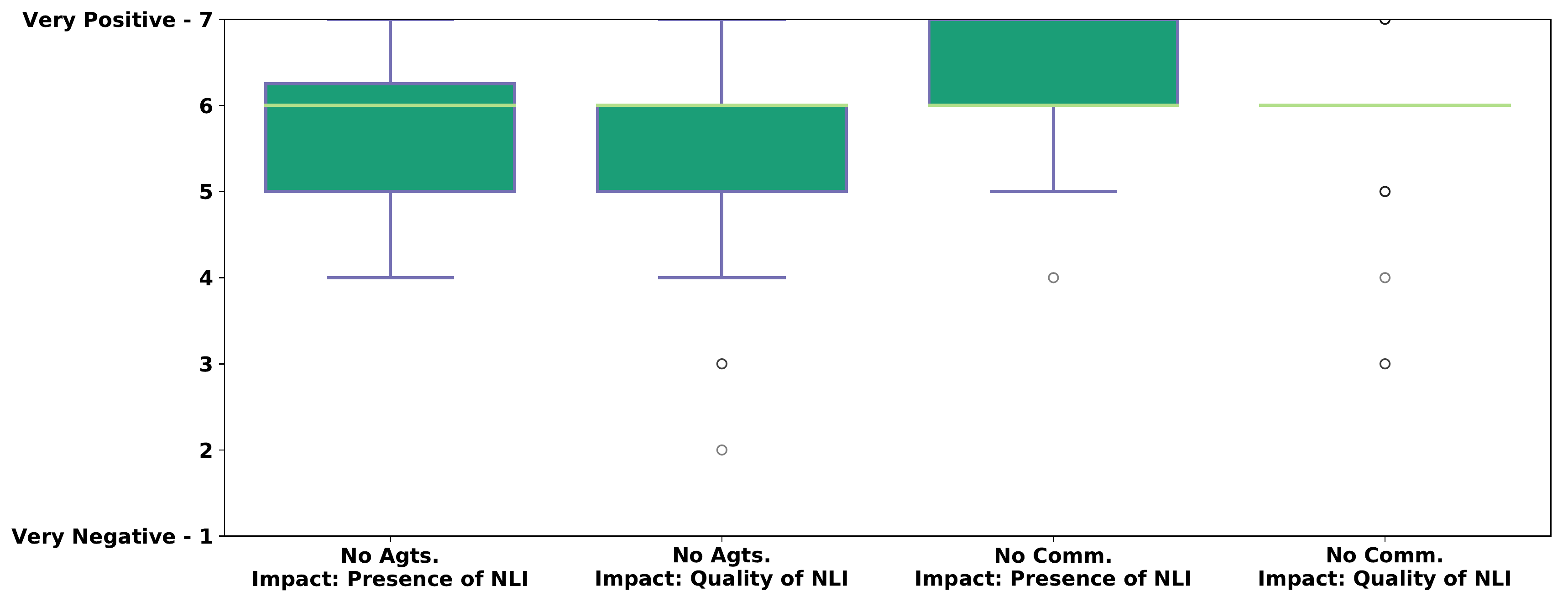}
	\caption{{\textbf{Participant Impact Perception:}} This figure shows perceived impacts of SPA over several factors. The presence and quality of natural language interactions produced by SPA significantly impact participant preferences. \textbf{No Agts.}: Compared against the no agents condition, participants perceived the presence and quality of natural language interactions (NLI) provided significant positive impacts on their preference ($5.82 \pm 0.94$, $5.29 \pm 1.24$). \textbf{No Comms.}: Compared against the no communication condition, the presence and quality of NLI was a more significant factor in preference ($6.18 \pm 0.77$, $5.75 \pm 1.00$). This is expected, as the presence of the virtual agents is less impactful in this case.}
	\label{nlpcrowds:fig:userStudy_plots_compare}
\end{figure*}

\begin{figure}[t]
	\centering
	\includegraphics[width=1\linewidth]{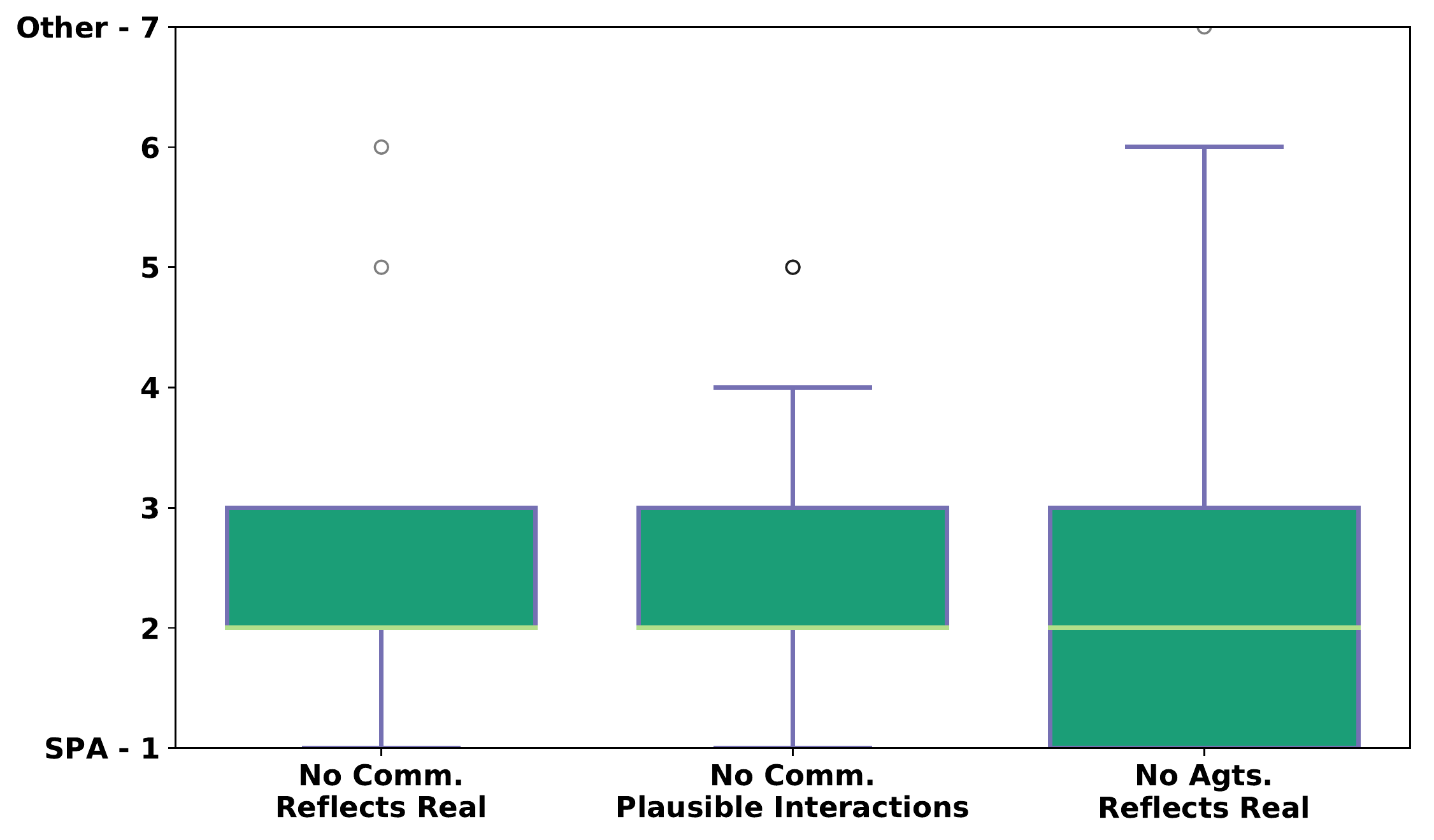} \\
	
	\caption{{\textbf{Participant Preference in User Evaluation}} This figure details participant preference for SPA compared to prior methods. The questions represented are ``Which simulation more closely reflects a real-world scenario`` and ``In which simulation did the interactions between the user and the agents seem more plausible``. Scores are normalized such that 1 indicates a strong preference for SPA and 7 indicates a strong preference against. In both conditions, no agents and no communication, participants rated SPA preferably in terms of generating more realistic simulations ($2.29 \pm 1.15$, $2.29 \pm 0.81$). In addition, participants felt the natural-language interactions increased the plausibility of the agent-avatar interactions compared to agents without the ability to interact using natural-language ($2.46 \pm 1.20$)}
	\label{nlpcrowds:fig:userStudy_plots_absolute}
\end{figure}

We conducted a user study to evaluate the plausibility of agent-avatar interactions and the overall simulation generated as a result of our algorithm.
Prior work establishes a procedure for evaluating new features of interactive agents against agents lacking the new capability\citep{nass2000truth}.
In addition, prior work has demonstrated that implausible behavior from agents can lead to a reduction in the sense of quality and presence in a virtual environment\citep{slater:2009:experience,Bailenson:2005:IIE}.
We therefore sought to establish whether our approach generated improvement in the overall perception of simulations between virtual agents and avatars compared with agents lacking the SPA interaction approach. We provide a summary of our user evaluation in this section and refer the reader to the supplemental document for extended details.

This study was conducted based on a within-subjects, paired-comparison design. 
Each scenario was displayed with a text-based prompt to provide the appropriate context. 
Participants were shown two pre-recorded videos of a subject interacting with the system in a side-by-side comparison of our method and one of two comparison methods, one with no virtual agents and one with agents lacking SPA.
The study consisted of two trials described in \cref{nlpcrowds:fig:userstudy_benchmarks}.
After each trial, participants were asked to answer a short questionnaire before moving on to the next scenario. 
The order of scenario and the positioning of the methods was counterbalanced.

Participants indicated responses on a Likert scale from 1 to 7.
Participants were asked to indicate which simulation more closely reflected a real-world scenario, and were asked the impact of the following items on their preference: the presence of natural language communication, the quality of the verbal responses from the agents, and the quality of the animation. 
The study was taken by 14 participants. 
For results reporting, we normalize participant responses such that 1 indicates a strong preference for our method for comparative questions and 7 indicates a strong positive impact for absolute questions.
We additionally collapsed participant responses across trials.

\textbf{Analysis:} We found statistical significance on all metrics.
For our analysis methods, we refer the reader to the supplemental material. 
The participant responses clearly demonstrate the benefits of our algorithm in terms of generating plausible agent-avatar interactions ($2.46 \pm 1.20$).
Expanding the interaction capacity of the virtual agents beyond movement interactions has a positive impact, and participants believed the quality of the natural-language interaction generated by SPA was a positive factor when comparing against agents without the ability to interact (mean impact:$5.75 \pm 1.00$) or cases with no agents (mean impact: $5.29 \pm 1.24$).
In both comparative conditions, either without agents or without communication, participants found our method to generate simulations and interactions with better reflect real-world scenarios ($2.29 \pm 1.15$ and $2.29 \pm 1.46$).

Overall, the results of our study show that participants find agents with our SPA natural-language interaction capability to be significantly more 
plausible than those without.
More importantly, they show that participants found the natural-language interactions generated by SPA to be a significantly positive factor on their preferences.
This indicates that SPA yields natural-language interactions which are plausible and effective. \Cref{nlpcrowds:fig:userStudy_plots_compare} and \cref{nlpcrowds:fig:userStudy_plots_absolute} provide further details of our analysis.
Our evaluation indicates that further work is merited to evaluate SPA against current proposed methods for multi-agent, multi-avatar interactions and establishes a baseline for comparison.

%% file: VR2020_arXiv/tex/conclusion.tex
\section{Conclusion}

We have presented a novel algorithm for generating virtual agent plans under uncertain conditions and natural language interactions between virtual agents and avatars for multi-agent multi-avatar environments.
Our approach allows agents to plan with uncertain information, engage in question and answer-based dialog and effectively accomplish their individual goals while facilitating plausible avatar-agent interactions.
We have demonstrated how our approach can be used to provide significant improvements to behavior plausibility for virtual agents in a shared environment and detailed a user-study which provides preliminary verification of our approach's advantage.
Moreover, our approach can simulate dozens of interactive virtual agents in real time.
Overall, SPA seeks to improve limitations of interactivity in typical multi-agent planning approaches and addresses limitations of single agent-avatar interactions in typical conversation agent approaches.

Our algorithm is part of ongoing research and has some limitations.
As with many propositional approaches, our algorithm's performance degrades with the complexity of the agents' desires and the problem domain. 
This could be addressed by planning as a pre-computation step and caching plan templates for subsequent simulations. 
We will additionally explore partial-order planning \cite{Russell:2009:AIM:1671238} to improve plan computation time. 

In addition, while the use of shallow-semantic parsing enables verbal interactions, it is very sensitive to training data, making the communication quality sensitive to lexicon quality.
Our agents are able to produce a relatively small set of dialog interactions. 
In the future, we will seek to expand the interaction capability of our virtual agents to include conversational context keeping and other dialog actions as well.
We will also explore how SPA can be integrated with existing context-aware dialogue approaches.
We will also work to improve our attention models. 
Research in human-agent interaction offers avenues for exploring different attention models \cite{Narang:2016:PedVR}.
Finally, our evaluation establishes that SPA generates plausible interactions.
Future work is needed to evaluate our algorithm against proposed state-of-the-art multi-agent multi-avatar interaction approaches.

%% file: VR2020_arXiv/tex/appendix_title.tex
\section{Appendix}

%% file: VR2020_arXiv/tex/appendix.tex
\appendix

\begin{figure*}[t]
	\centering
	\includegraphics[width=1\linewidth]{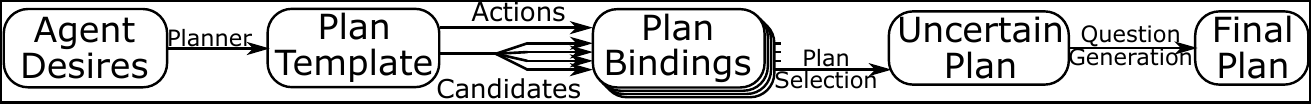} 
	
	\caption{\label{nlpcrowds:fig:nlp_planner_pipe} \textbf{Two-stage Action Planning with Incomplete Information}: Each agent is given a desire to achieve during simulation. We propose a two-stage planner which generates action plans despite uncertainties in the agent's knowledge. The first stage generates a plan template and a set of candidates for each argument in the plan. The second stage generates a set of candidate bindings. The algorithm selects the plan with the least uncertainty and generates an action plan from the bindings. If any uncertain information is present, an uncertainty resolution action is created, yielding the final action plan which may include asking questions, or exploring the environment.}
	
\end{figure*}

\begin{figure*}[t]
	\centering
	\includegraphics[width=1\linewidth]{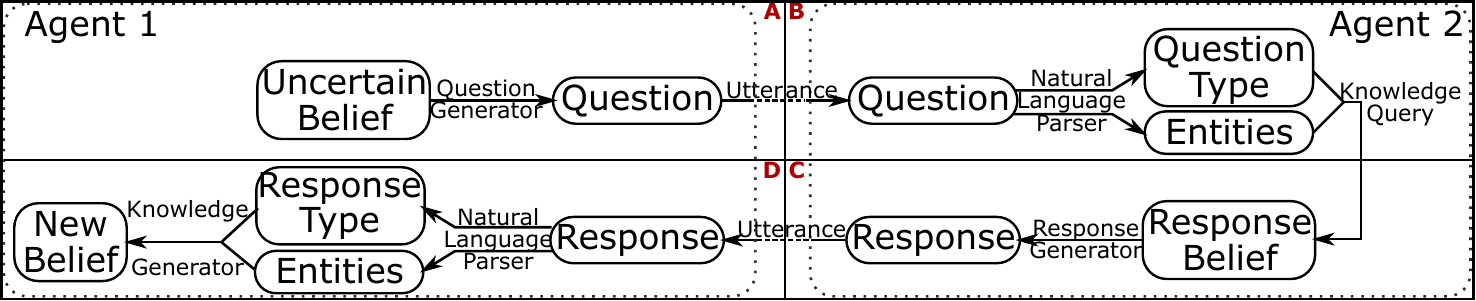} 
	
	\caption{\label{nlpcrowds:fig:nlp_conversation_pipe} \textbf{Natural Language Communication for Virtual Agents}: This figure illustrates a sample interaction between two agents using our natural-language approach (clockwise from top). \textbf{(A)}~Agent 1's plan yields an uncertain belief.  The agent generates a question from the belief template. The question is communicated as a natural language utterance. \textbf{(B)}~Agent 2 receives the utterance and parses it into the relevant question type and entities. The agent queries its knowledge-base for an answer to the question, yielding a response predicate. \textbf{(C)}~Agent 2 uses our approach to generate a response utterance. \textbf{(D)}~Agent 1 receives the utterance, generates the appropriate response type and entities, and processes these into a new belief which is stored in the knowledge-base.}
\end{figure*}

\section{Additional Method Details}

\Cref{nlpcrowds:fig:nlp_planner_pipe} details our two-stage planning pipeline. 
\Cref{nlpcrowds:fig:nlp_conversation_pipe} details our natural-language communication pipeline in an example dialog.

\textbf{Sample Lexion Entry:} The lexicon used to generate the results presented in section 6 of the main document consists of a small subset of English words annotated with sample sentences and reference hints for the parser.
The word ``location'' as it appears in the lexicon is tagged with the hint ``where'' and ``InSpace'' which are other forms seen in our domain descriptions.
As a predicate, It is assigned several template sentences with annotated natural-language intentions.
One such sentence with the label \emph{predicate answer} is 

``the [PREDICATE:NAME] of [PREDICATE-ENTITY:DEF-ARTICLE-NAME] is [PREDICATE-ENTITY:DEF-ARTICLE-NAME:GALLERY]. ''

This template provides parameters for binding an entity with a definite article, e.g. the statue, to an entity with the type specifier gallery. A sample binding of the sentence would be 

``The location of the Venus de Milo is Gallery B''.

\section{Additional Performance Results}

\subsection{Implementation and Performance Benchmarks}

Our experiments were conducted on a desktop pc with an Intel Xeon E5 CPU, NVIDIA TitanX GPU and 16gb of RAM. We coupled our propositional planner with Rasa NLU ~\cite{rasanlu} for semantic parsing. User utterances were captured via microphone and automated speech recognition. Our algorithm was implemented in python, and our VR experiments were performed with the Occulus Rift HMD. We couple our approach with the 3D animation system described in \cite{narang:2018:FBCrowd}.

In addition to the results reported in section 6 of the main document, we evaluated the algorithm's performance as a function of domain size and number of agents. Consistent with prior propositional planning approaches, our algorithm scales linearly in the number of agents and exponentially in the size of the problem domain. \Cref{nlpcrowds:fig:performance_macro_graphs} details our experimental results. \Cref{nlpcrowds:table:benchmark_details} provides additional details about the number of agents, desires, and verbal interactions in our benchmarks.

\input{SIG19_arXiv/tex/user_study_full.tex}

%% file: SIG19_arXiv/tex/user_study_full.tex
\subsection{User Evaluation}

This section provides a formal description of the user study we conducted to evaluate the plausibility of agent-avatar interactions and the overall simulation generated as a result of our algorithm. In addition, we provide the complete set of participant responses and additional response analysis.

\textbf{Experiment Goals \& Expectations:} We hypothesize that verbal communication between agents and avatars will enhance the perceived plausibility of the simulation, and generate positive impressions as compared to the control conditions. 

\subsubsection{Comparison Conditions}

\begin{itemize}
	\item \textbf{No Agents}: In the no agents case, a user avatar explores a virtual environment without any virtual agents present.
	\item \textbf{No Communication}: In the no communication case, a user avatar explores a virtual environment with agents who could not interact using natural-language communication.
\end{itemize}
	
\subsubsection{Experimental Design}

 This study was conducted based on a within-subjects, paired-comparison design. 
 Each scenario was displayed with a text-based prompt to provide the appropriate context. 
 Participants were shown two pre-recorded videos of a subject interacting with the system in a side-by-side comparison of our method and one of the comparison methods. 
 They were then asked to answer a short questionnaire before moving on to the next scenario. 
 The order of scenario and the positioning of the methods was counterbalanced.

\subsubsection{Environments}

The multi-agent tradeshow scenario and multi-agent museum were used in this study.
Three confederates were recruited to participate as the avatar in the environments. 
In trials using our method, the confederate was allowed to interact with the agents using natural-language communication.
In each case, the avatar was piloted from a first-person view. Their interactions were recorded via screen capture and a microphone.

\textbf{Tradeshow}: The avatar was instructed to find the ``registration booth''. They were shown a picture of the booth before beginning their task but were not told its location. In the SPA case, virtual agents in the environment were able to interact and provide the location of the booth to the avatar. We refer the reader to the main document for visual examples of the benchmarks.

\textbf{Museum}: The avatar was instructed to find a specific statue in the museum but was not told the location of the statue. 
The statue in question was Lucy, courtesy of the Stanford University Computer Graphics Laboratory.
In the SPA case, a virtual agent near the avatar's starting position was provided knowledge of the location. 
The avatar was able to ask this agent the location of the statue. 
In addition, two agents were placed along the path to the goal who would interrupt the avatar's progress and ask the avatar for the locations of other statues as they passed.

\subsubsection{Metrics}

Participants were asked a set of common questions for both comparison methods, with specific additions for each comparison method.

\textbf{Common Metrics}: Participants were asked to indicate which simulation more closely reflected a real-world scenario on a Likert scale with 1 indicating strong preference for the method presented on the left, 7 indicating strong preference for the method presented on the right, and 4 indicating no preference. They were then asked the impact of the following items on their preference: the presence of natural language communication, the quality of the verbal responses from the agents, and the quality of the animation. These were answered on a Likert scale with 1 indicating strong negative impact, 7 indicating strong positive impact, and 4 indicating no impact.

\textbf{No Agent Metrics}: Participants were additionally asked what impact the presence of the virtual agents had on their preference. 

\textbf{No Communication Metrics}: Participants were additionally asked which of the methods demonstrated more plausible interactions, in which simulation did the agents appear to benefit more from their interactions with the avatar, and in which simulation did the avatar appear to benefit more from their interactions with the virtual agents.

\subsubsection{Results}

The study was taken by 14 participants. 
We normalized the data for comparative questions such that a response of $1$ indicates strong preference for our method.
We collapsed the common metrics across trials as well as plausibility of interactions question for the No Communication metric and the presence of virtual agents from the No Agents metric.
We performed a one-sample t-test comparing the mean of each question with a hypothetical mean of 4 (no preference or no impact).
We limit our discussion below to questions which directly deal with natural-language interaction and preference for the methods. \Cref{nlpcrowds:table:userstudy_frequency} gives complete details of the participant responses collected for our user evaluation. 

\begin{table*}[t]
	\centering
	\caption{{\bf Performance Benchmark Details}. We detail number of agents, desires, and the NL-I details of the benchmark scenarios including how many questions were asked, statements made, facts overheard by agents, and parser failures. We observe, as expected, that as the number of agents and desires increases, the amount of information gained from overhearing nearby agents increases.}
	\label{nlpcrowds:table:benchmark_details}
	
	\begin{tabular}{llrrrrrr}
		\toprule
		Scene &  Agents &  Desires &  Statements & Questions & Facts Overheard &  
		Parser failures \\
		\midrule
		Anti-podal Circle &    10 &    30 &   14 &   5 &      28  &  0 \\
		Evacuation &  11 &   10 &   4 &  0 &   20 &  0 \\
		Museum &    5 &  9 &   20 &     13 &      3 &      0 \\
		Trade Show &    4 &  1 &    3 &   2 &      0 &      0 \\
		\bottomrule
	\end{tabular}
\end{table*}

We found the question ``Which simulation more closely reflects a real-world scenario'' significant in both the no agents condition 
($t(27)=-6.204, p < 0.000$), 
and the no communication condition 
($t(27)=-7.887, p < 0.000$).
We found the question ``What impact did the presence of natural language interaction have on your answer'' significant in both the no agents condition ($t(27)=10.200, p < 0.000$),
and the no communication condition 
($t(27)=14.925, p < 0.000$).
We found the question ``What impact did the quality of the verbal responses from the agents have on your answer'' significant in both the no agents condition 
($t(27)=5.473, p < 0.000$), 
and the no communication condition
($t(27)=9.218, p < 0.000$).
We found the question ``In which simulation did the interactions between the user and the agents seem more plausible'' significant in the no communication condition 
($t(27)=-6.765, p < 0.000$). 
It was not asked of the no agents condition.
We found the question ``What impact did the presence of virtual agents have on your answer'' significant in the no agents condition 
($t(27)=13.478, p < 0.000$). 
It was not asked of the no communication condition.
perception of the impact of natural language interactions.

\textbf{Analysis:} As detailed in section 6 of the main document, participant responses demonstrate the benefits of our algorithm in terms of generating plausible agent-avatar interactions ($2.46 \pm 1.20$).
In both comparative conditions, either without agents or without communication, participants found our method to generate simulations and interactions with better reflect real-world scenarios ($2.29 \pm 1.15$ and $2.29 \pm 1.46$).

\begin{figure*}[t]
	\centering
	\begin{tabular}{cc}
		\includegraphics[width=0.45\textwidth]{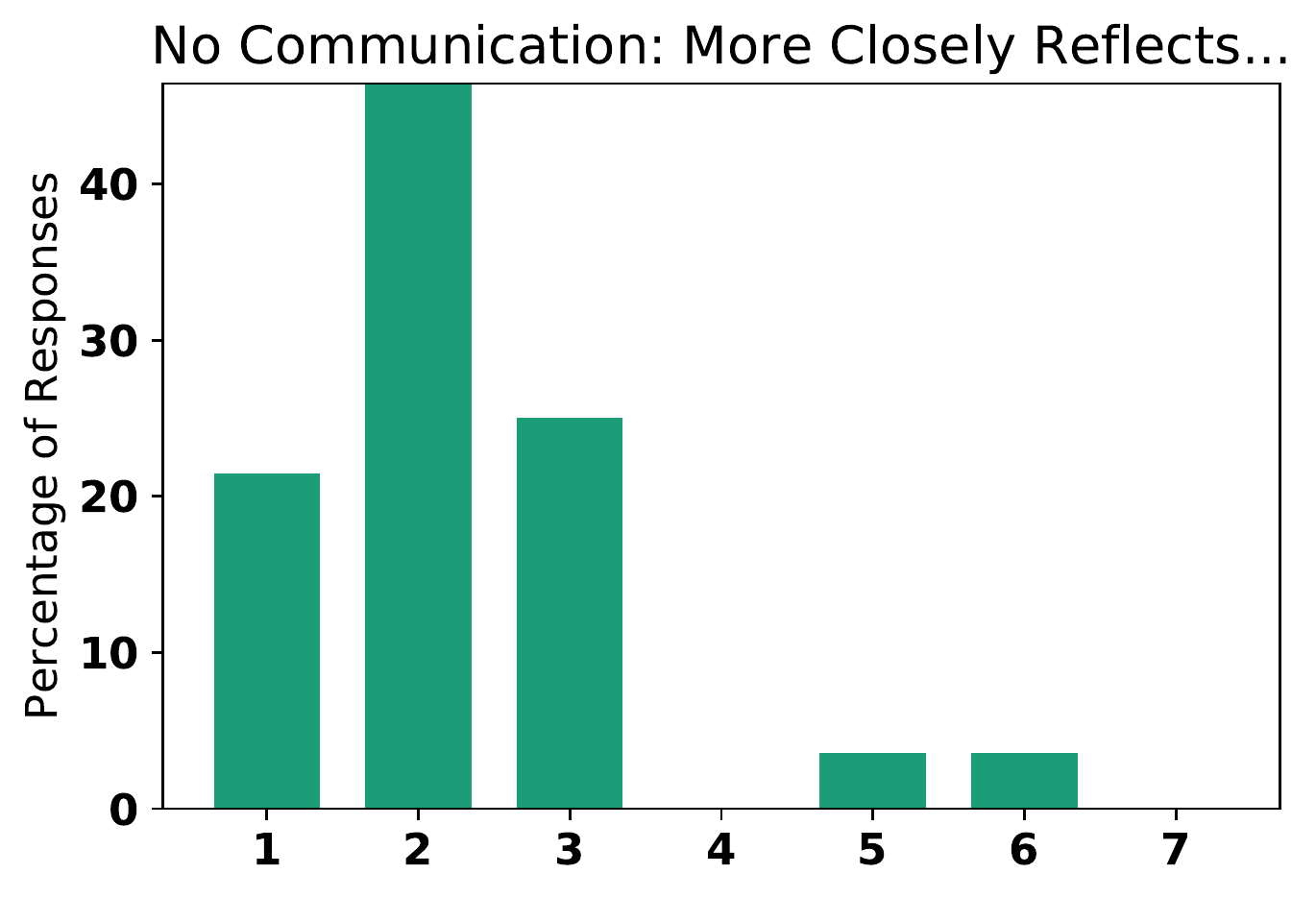} &
		\includegraphics[width=0.45\textwidth]{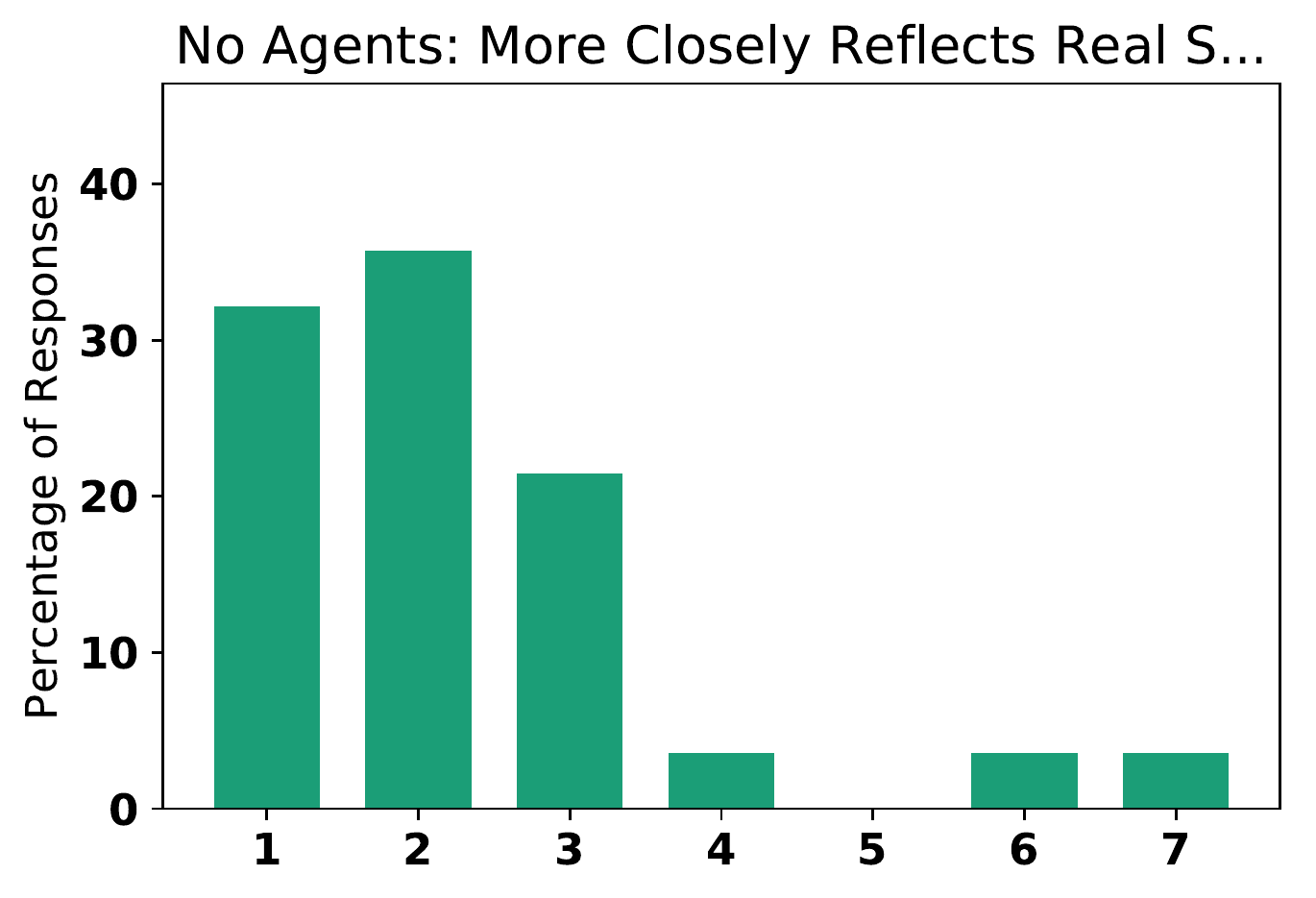}
		\\
		\textbf{(A)} & \textbf{(B)} 
	\end{tabular}
	\caption{\textbf{Histogram data of user responses for Which scenario better reflects real-world scenarios}: Participants in our evaluation found simulations using SPA significantly more plausible compared to simulations with a prior approach {\bf (A)} and simulations with no agents {\bf (B)}. This indicates that the presence of agents has a positive impact on plausibility and that our agents behave sufficiently well to increase plausibility with respect to agents lacking SPA. }
	\label{nlpcrowds:fig:histogram_preferences_realworld}
\end{figure*}

\begin{figure*}[t]
	\centering
	\begin{tabular}{cc}
		\includegraphics[width=0.45\textwidth]{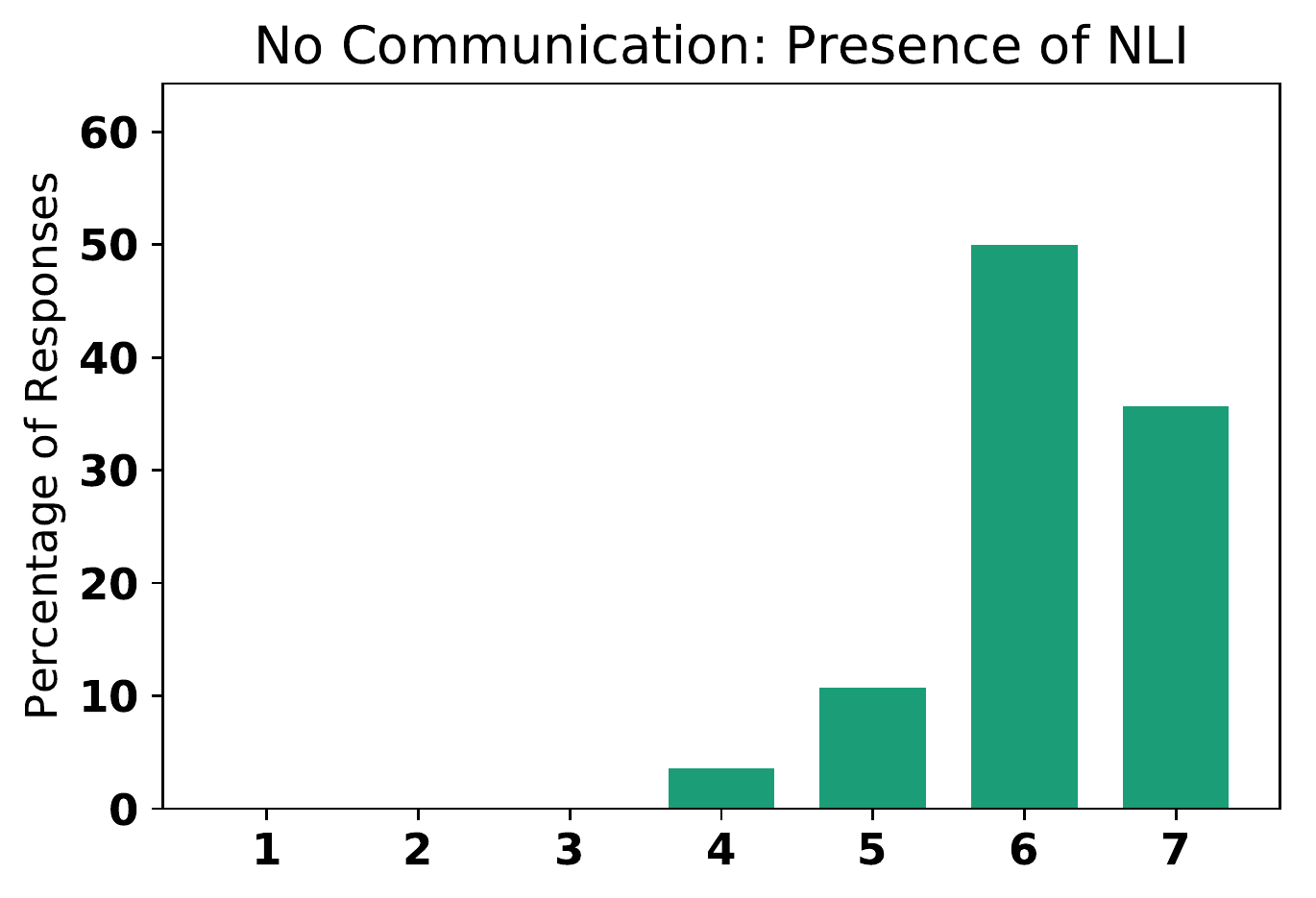} &
		\includegraphics[width=0.45\textwidth]{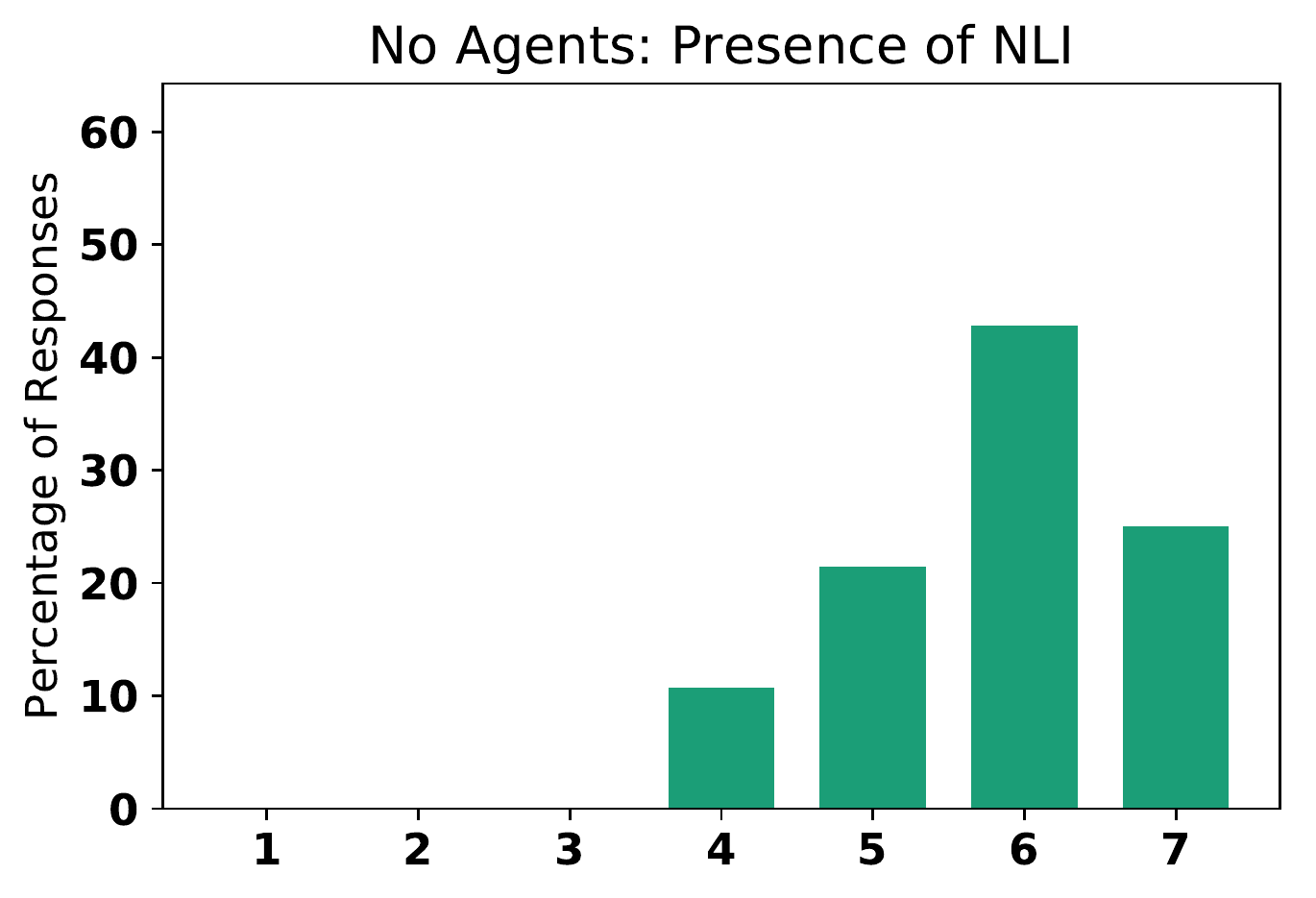}
		\\
		\textbf{(A)} & \textbf{(B)}
	\end{tabular}
	\caption{\textbf{Histogram data of user responses for impact of natural language interactions} Participants in our evaluation found the presence and quality of the natural language interactions had a significant impact on their preference for our approach to simulations with agents lacking SPA {\bf A} and simulations without agents {\bf b}. In addition, the preference for the quality of the natural language interactions generated with SPA is stronger when compared to agents not able to communicate.}
	\label{nlpcrowds:fig:histogram_preferences_nli}
	\vspace*{-0.1in}
\end{figure*}

Overall, Participants preferred our approach in 84\% of responses. 
Of those responses, 84\% were strong preferences ($r \leq 2$).
\Cref{nlpcrowds:fig:histogram_preferences_realworld} and \Cref{nlpcrowds:fig:histogram_preferences_nli} provide additional details about our method's advantages over prior approaches.

\begin{figure*}[t]
	\centering
	\begin{tabular}{cc}
	\includegraphics[width=0.49\textwidth]{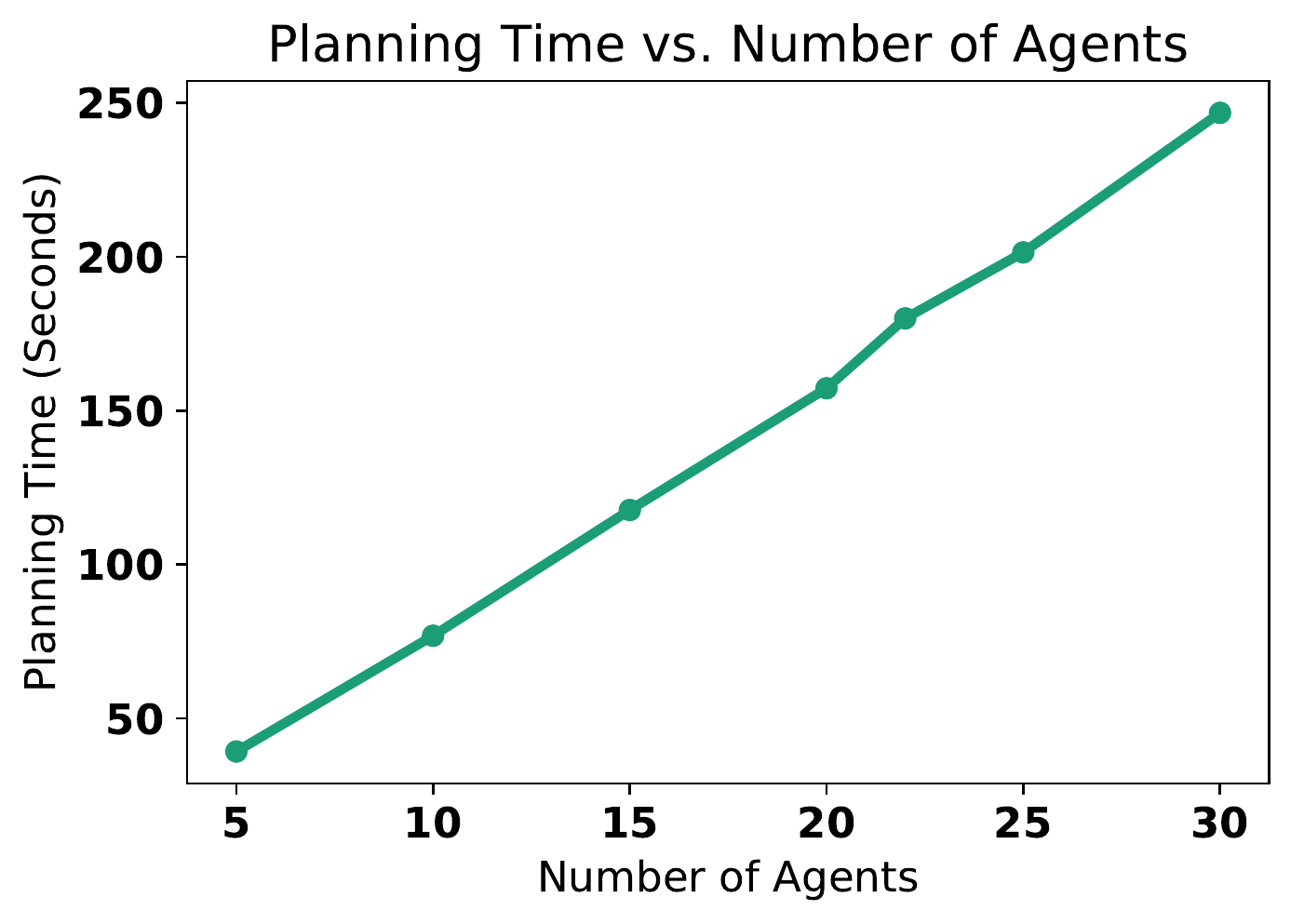} &
	\includegraphics[width=0.51\textwidth]{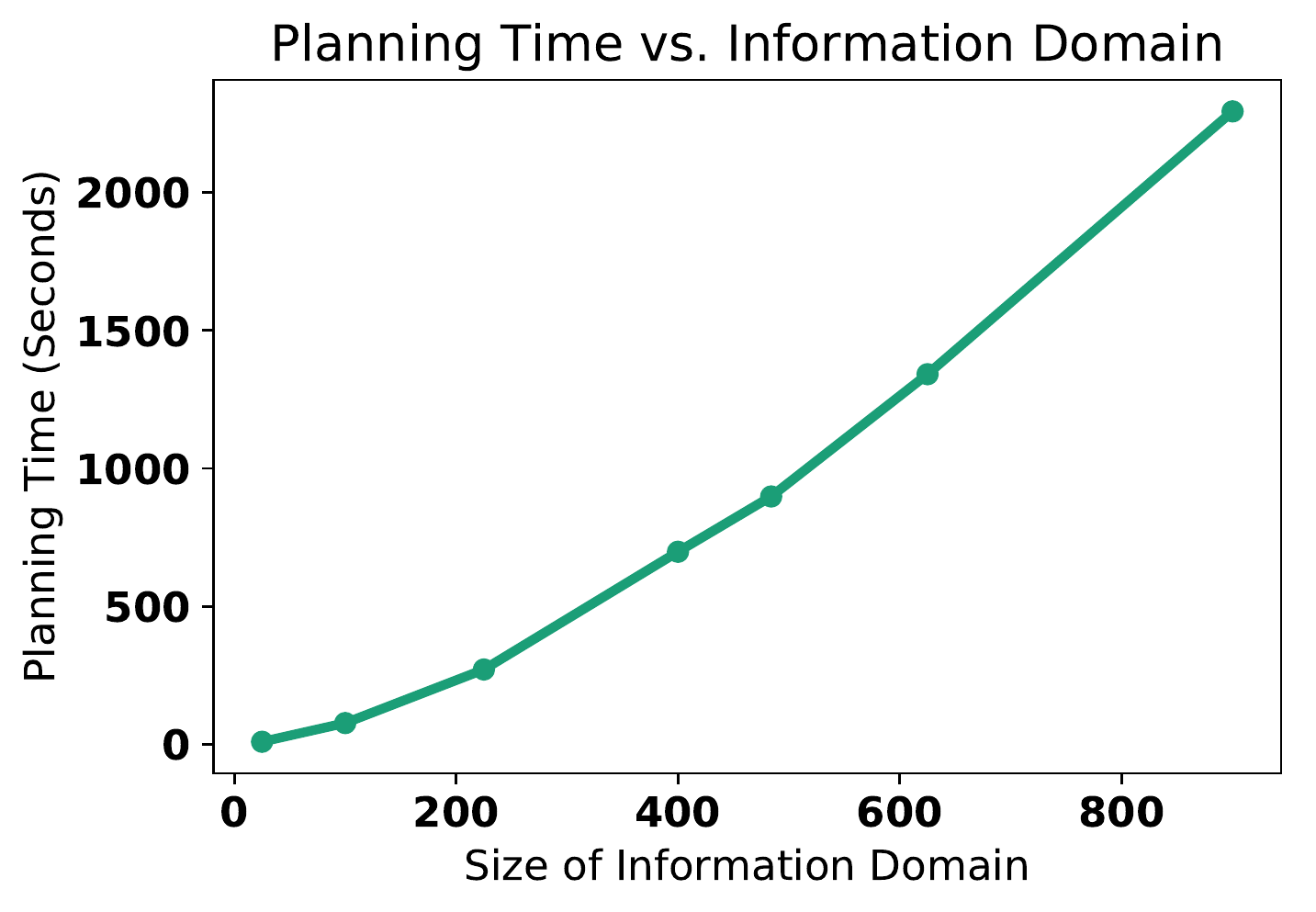} \\
	(A) & (B) 
	\end{tabular}
	\caption{\textbf{Performance Results Varying Number of Agents and Problem Size}: \textbf{(A) Varying agents on a fixed domain size (100 predicates):} We observe that our algorithmic approach's performance scales linearly in the number of agents. \textbf{(B) Varying domain size for a fixed number of agents (10 agents):} We observe that our algorithm's performance scales exponentially in the size of the problem domain. This is consistent with propositional approaches. However, our two stage planning approach enables rapid replanning after the initial planning step, reducing overall planning time.}
	\label{nlpcrowds:fig:performance_macro_graphs}
	\vspace*{-0.1in}
\end{figure*}

\begin{table*}[h]
	\centering
	\caption{{\bf Frequency of Responses in User Evaluation}. This table shows the frequency of participant responses in the user evaluation, as well as the means and p-value for a one-sample t-test with a hypothetical mean of 4. For comparative questions, responses less than 4 indicate preference for our agents. For impact questions, responses greater than 4 indicate positive impacts. We found participant responses to all question significant.}
	\label{nlpcrowds:table:userstudy_frequency}
	\begin{tabular}{lrrrrrrrrrr}
		\toprule
		Question & 1 & 2 & 3 & 4 & 5 & 6 & 7 & mean & std & p-value \\
		\midrule
		\multicolumn{11}{l}{{\bf NL-I Agents vs Non-Interactive Agents }}\\
		\multicolumn{11}{l}{{\bf Comparative Questions (NL-I Agents left) }}\\
		\midrule
		More closely reflects real scenario & 6 & 13 & 7 & 0 & 1 & 1 & 0 & 2.29 & $\pm 1.15$ & $< 0.000$ \\
		Agents benefit more from interaction & 11 & 4 & 1 & 11 & 0 & 1 & 0 & 2.57 & $\pm 1.53$ & $ < 0.000 $ \\
		User benefits more from interaction & 17 & 10 & 0 & 0 & 0 & 1 & 0 & 1.54 & $\pm 1.00$ & $ < 0.000 $ \\
		More plausible interactions & 5 & 13 & 5 & 2 & 3 & 0 & 0 & 2.46 & $\pm 1.20$ & $ < 0.000 $ \\
		\multicolumn{11}{l}{{\bf Impact Questions }}\\
		Presence of natural Language & 0 & 0 & 0 & 1 & 3 & 14 & 10 & 6.18 & $\pm 0.77$ & $ < 0.000 $ \\
		Quality of the verbal interactions & 0 & 0 & 2 & 1 & 3 & 18 & 4 & 5.75 & $\pm 1.00$ & $ < 0.000 $ \\
		Animation of the virtual agents & 0 & 0 & 4 & 13 & 3 & 3 & 5 & 4.74 & $\pm 1.36$ & 0.010 \\
		\midrule
		\multicolumn{11}{l}{{\bf NL-I Agents vs No Agents }}\\
		\multicolumn{11}{l}{{\bf Comparative Questions (NL-I Agents left) }}\\
		\midrule
		More closely reflects real scenario & 9 & 10 & 6 & 1 & 0 & 1 & 1 & 2.29 & $\pm 1.46$ & $ < 0.000 $ \\
		\midrule
		\multicolumn{11}{l}{{\bf Impact Questions }}\\
		\midrule
		Presence of the virtual agents & 0 & 0 & 0 & 0 & 8 & 10 & 10 & 6.07 & $\pm 0.81$ & $ < 0.000 $ \\
		Actions of the virtual agents & 0 & 0 & 0 & 6 & 9 & 10 & 3 & 5.36 & $\pm 0.95$ & $ < 0.000 $ \\
		Presence of natural Language & 0 & 0 & 0 & 3 & 6 & 12 & 7 & 5.82 & $\pm 0.94$ & $ < 0.000 $ \\
		Quality of the verbal interactions & 0 & 1 & 2 & 3 & 7 & 12 & 3 & 5.29 & $\pm 1.24$ & $ < 0.000 $ \\
		Animation of the virtual agents & 0 & 0 & 3 & 11 & 10 & 2 & 2 & 4.61 & $\pm 1.03$ & 0.004 \\
		\bottomrule
	\end{tabular}
\end{table*}

%% file: VR2020_arXiv/tex/acknowledgments.tex
\acknowledgments{
This work was supported in part by ARO Grants W911NF1910069 and W911NF1910315 and Intel.
}